\documentclass[10pt,journal,compsoc]{IEEEtran}
\ifCLASSOPTIONcompsoc
  \usepackage[nocompress]{cite}
\else
  \usepackage{cite}
\fi
\ifCLASSINFOpdf
  \usepackage[pdftex]{graphicx}
  \graphicspath{{../pdf/}{../jpeg/}}
\else
  \usepackage[dvips]{graphicx}
  \graphicspath{{../eps/}}
\fi
\usepackage{amsmath}
\usepackage[table,xcdraw]{xcolor}
\usepackage{caption}
\usepackage{subcaption}
\usepackage{hyperref}
\usepackage{multirow}
\usepackage[table,xcdraw]{xcolor}
\usepackage{xspace}
\usepackage{multirow}
\usepackage[table,xcdraw]{xcolor}
\usepackage{tabularx}
\usepackage{fontawesome}
\usepackage{ragged2e}
\usepackage[inline]{enumitem}

\newcommand{\sys}{\texttt{TimeLighting}\xspace}

\begin{document}

%% Paper title.
\title{TimeLighting: %Unveiling New Dimensions in 
Guided Exploration of 2D Temporal Network Projections}

%% This is how authors are specified in the journal style

%
%
% author names and IEEE memberships
% note positions of commas and nonbreaking spaces ( ~ ) LaTeX will not break
% a structure at a ~ so this keeps an author's name from being broken across
% two lines.
% use \thanks{} to gain access to the first footnote area
% a separate \thanks must be used for each paragraph as LaTeX2e's \thanks
% was not built to handle multiple paragraphs
%
%
%\IEEEcompsocitemizethanks is a special \thanks that produces the bulleted
% lists the Computer Society journals use for "first footnote" author
% affiliations. Use \IEEEcompsocthanksitem which works much like \item
% for each affiliation group. When not in compsoc mode,
% \IEEEcompsocitemizethanks becomes like \thanks and
% \IEEEcompsocthanksitem becomes a line break with idention. This
% facilitates dual compilation, although admittedly the differences in the
% desired content of \author between the different types of papers makes a
% one-size-fits-all approach a daunting prospect. For instance, compsoc 
% journal papers have the author affiliations above the "Manuscript
% received ..."  text while in non-compsoc journals this is reversed. Sigh.

\author{Velitchko Filipov$^{1}$, 
        Davide Ceneda$^{1}$,
        Daniel Archambault$^{2}$,
        Alessio Arleo$^{1,3}$%
        
\IEEEcompsocitemizethanks{
    \IEEEcompsocthanksitem 
    $^{1}$ Centre for Visual Analytics Science and Technology (CVAST), TU Wien, Vienna (AT). E-mail: \{name.surname\}@tuwien.ac.at.\protect\\
     $^{2}$ School of Computing, Newcastle University, Newcastle (UK). E-mail: \ daniel.archambault@newcastle.ac.uk.\protect\\
      $^{3}$ Visualization Cluster, Eindhoven University of Technology, Eindhoven (NL). E-mail: \ a.arleo@tue.nl.\protect\\
    % E-mail: \{name.surname\}@tuwien.ac.at\protect
    \IEEEcompsocthanksitem A preliminary version of this research has been presented at the International Symposium on Graph Drawing and Network Visualization (GD) 2023~\cite{filipov2023gd}, selected by the Program Chairs, and invited for publication in TVCG. Part of the text has been reused in this publication.\protect\\

}
}

% note the % following the last \IEEEmembership and also \thanks - 
% these prevent an unwanted space from occurring between the last author name
% and the end of the author line. i.e., if you had this:
% 
% \author{....lastname \thanks{...} \thanks{...} }
%                     ^------------^------------^----Do not want these spaces!
%
% a space would be appended to the last name and could cause every name on that
% line to be shifted left slightly. This is one of those "LaTeX things". For
% instance, "\textbf{A} \textbf{B}" will typeset as "A B" not "AB". To get
% "AB" then you have to do: "\textbf{A}\textbf{B}"
% \thanks is no different in this regard, so shield the last } of each \thanks
% that ends a line with a % and do not let a space in before the next \thanks.
% Spaces after \IEEEmembership other than the last one are OK (and needed) as
% you are supposed to have spaces between the names. For what it is worth,
% this is a minor point as most people would not even notice if the said evil
% space somehow managed to creep in.

% The paper headers
\markboth{Journal of \LaTeX\ Class Files,~Vol.~14, No.~8, August~2015}%
{Filipov \MakeLowercase{\textit{et al.}}: TimeLighting: Guided Exploration of 2D Temporal Network Projections}
% The only time the second header will appear is for the odd numbered pages
% after the title page when using the twoside option.
% 
% *** Note that you probably will NOT want to include the author's ***
% *** name in the headers of peer review papers.                   ***
% You can use \ifCLASSOPTIONpeerreview for conditional compilation here if
% you desire.

% The publisher's ID mark at the bottom of the page is less important with
% Computer Society journal papers as those publications place the marks
% outside of the main text columns and, therefore, unlike regular IEEE
% journals, the available text space is not reduced by their presence.
% If you want to put a publisher's ID mark on the page you can do it like
% this:
%\IEEEpubid{0000--0000/00\$00.00~\copyright~2015 IEEE}
% or like this to get the Computer Society new two part style.
%\IEEEpubid{\makebox[\columnwidth]{\hfill 0000--0000/00/\$00.00~\copyright~2015 IEEE}%
%\hspace{\columnsep}\makebox[\columnwidth]{Published by the IEEE Computer Society\hfill}}
% Remember, if you use this you must call \IEEEpubidadjcol in the second
% column for its text to clear the IEEEpubid mark (Computer Society jorunal
% papers don't need this extra clearance.)

% use for special paper notices
%\IEEEspecialpapernotice{(Invited Paper)}

% for Computer Society papers, we must declare the abstract and index terms
% PRIOR to the title within the \IEEEtitleabstractindextext IEEEtran
% command as these need to go into the title area created by \maketitle.
% As a general rule, do not put math, special symbols or citations
% in the abstract or keywords.
\IEEEtitleabstractindextext{%
\begin{abstract}
\justifying
In temporal (\textit{event-based}) networks, time is a continuous axis, with real-valued time coordinates for each node and edge. Computing a layout for such graphs means embedding the node trajectories and edge surfaces over time in a $2D + t$ space, known as the space-time cube. Currently, these space-time cube layouts are visualized through animation or by slicing the cube at regular intervals. However, both techniques present problems such as below-average performance on tasks as well as loss of precision and difficulties in selecting timeslice intervals. In this paper, we present \sys, a novel visual analytics approach to visualize and explore temporal graphs embedded in the space-time cube. Our interactive approach highlights node trajectories and their movement over time, visualizes node ``aging'', and provides guidance to support users during exploration by indicating interesting time intervals (``when'') and network elements (``where'') are located for a detail-oriented investigation. This combined focus helps to gain deeper insights into the temporal network's underlying behavior. We assess the utility and efficacy of our approach through two case studies and qualitative expert evaluation. The results demonstrate how \sys supports identifying temporal patterns, extracting insights from nodes with high activity, and guiding the exploration and analysis process.  

\end{abstract}

% Note that keywords are not normally used for peerreview papers.
}

% %other entries to be set up for journal
% \shortauthortitle{Filipov \MakeLowercase{\textit{et al.}}: XXXX}
% %\shortauthortitle{Firstauthor \MakeLowercase{\textit{et al.}}: Paper Title}
\maketitle
% To allow for easy dual compilation without having to reenter the
% abstract/keywords data, the \IEEEtitleabstractindextext text will
% not be used in maketitle, but will appear (i.e., to be "transported")
% here as \IEEEdisplaynontitleabstractindextext when the compsoc 
% or transmag modes are not selected <OR> if conference mode is selected 
% - because all conference papers position the abstract like regular
% papers do.
\IEEEdisplaynontitleabstractindextext
% \IEEEdisplaynontitleabstractindextext has no effect when using
% compsoc or transmag under a non-conference mode.

% For peer review papers, you can put extra information on the cover
% page as needed:
% \ifCLASSOPTIONpeerreview
% \begin{center} \bfseries EDICS Category: 3-BBND \end{center}
% \fi
%
% For peerreview papers, this IEEEtran command inserts a page break and
% creates the second title. It will be ignored for other modes.
\IEEEpeerreviewmaketitle

\IEEEraisesectionheading{\section{Introduction}\label{sec:introduction}} 
\IEEEPARstart{T}{emporal} (or \textit{event-based}) networks~\cite{12Holme} are dynamic graphs where the temporal dynamics, such as node and edge additions and removals, have real-time valued coordinates. These have been characterized and studied extensively~\cite{filipov2023we}, as they are used in many applications to model phenomena of commercial and academic interest, such as interactions in social media~\cite{12Holme}, communication networks~\cite{realitymining}, and contact tracing~\cite{emergencyresponse}, to name a few. In contrast to traditional dynamic graph drawing~\cite{beck2017,filipov2023we}, where time is discretized (or \textit{timesliced}), creating a visualization for temporal networks poses different challenges. Juxtaposition (or small multiples) presents the discrete timeslices in a side-by-side manner and requires identifying suitable timeslices, which inevitably leads to quantization errors. Such quantization errors obscure the fine temporal details that might be crucial in some domains (e.g., the exact order of personal contacts in contact tracing networks). 
Animation would not suffer from such artifacts, and it has been used in previous work on event-based graph drawing as a visualization metaphor to display the computed layouts~\cite{arleo2022event,simonetto2018event}.
Animation is a more natural way to encode time; however, it is not perceptually effective for many tasks involving dynamic networks~\cite{10FarrugiaIVJ,archambault2011}. In particular, animation falls short when comparing different timeslices (or states of the network) and the changes between distant timeslices (i.e., the first and last time slices) as it requires the viewer to memorize these.  
Moreover, the vast majority of research on animation has been done with timesliced graphs (see, e.g.,~\cite{archambault2011,archambault16,beck2017,10FarrugiaIVJ}), and its application to temporal networks is still in its infancy.

Temporal networks can also be drawn in a 3D space, namely, the space-time cube ($2D + t$)~\cite{hagerstrand1970people,17Bach}. In this case, a drawing algorithm computes the node trajectories over time and space. Existing research~\cite{arleo2022event,17SimonettoGD,simonetto2018event} provides evidence that this drawing approach yields better quality drawings of temporal graphs, compared to their timesliced counterparts (e.g., Visone~\cite{baur2001visone}), and for discrete-time graphs when many changes occur between timeslices. Despite this, research on visually depicting these trajectories and obtaining insights from their behavior is still an under-investigated topic. %- a gap that we intend to address in this paper. 

On these premises, we present \sys, a guidance-enhanced Visual Analytics (VA) solution for exploring node trajectories in the space-time cube keeping the \textit{full} temporal resolution of the network. Our approach supports understanding temporal patterns and behaviors and, in general, extracting insights from datasets with complex temporal dynamics. \sys has several key differences compared to existing methods. Unlike traditional dynamic graph drawing techniques that discretize time and often result in quantization errors, our approach maintains the full temporal resolution, preserving fine temporal details crucial in domains like contact tracing. Additionally, we integrate multiple elements to guide and ease the exploration process, highlighting key intervals for further inspection. We further present a movement score, based on the length of each node's trajectory over time that helps to rank and identify the more and less stable parts of the graph, facilitating a better understanding of the nodes' dynamics.

\begin{figure}[t!]
    \centering\includegraphics[width=\linewidth]{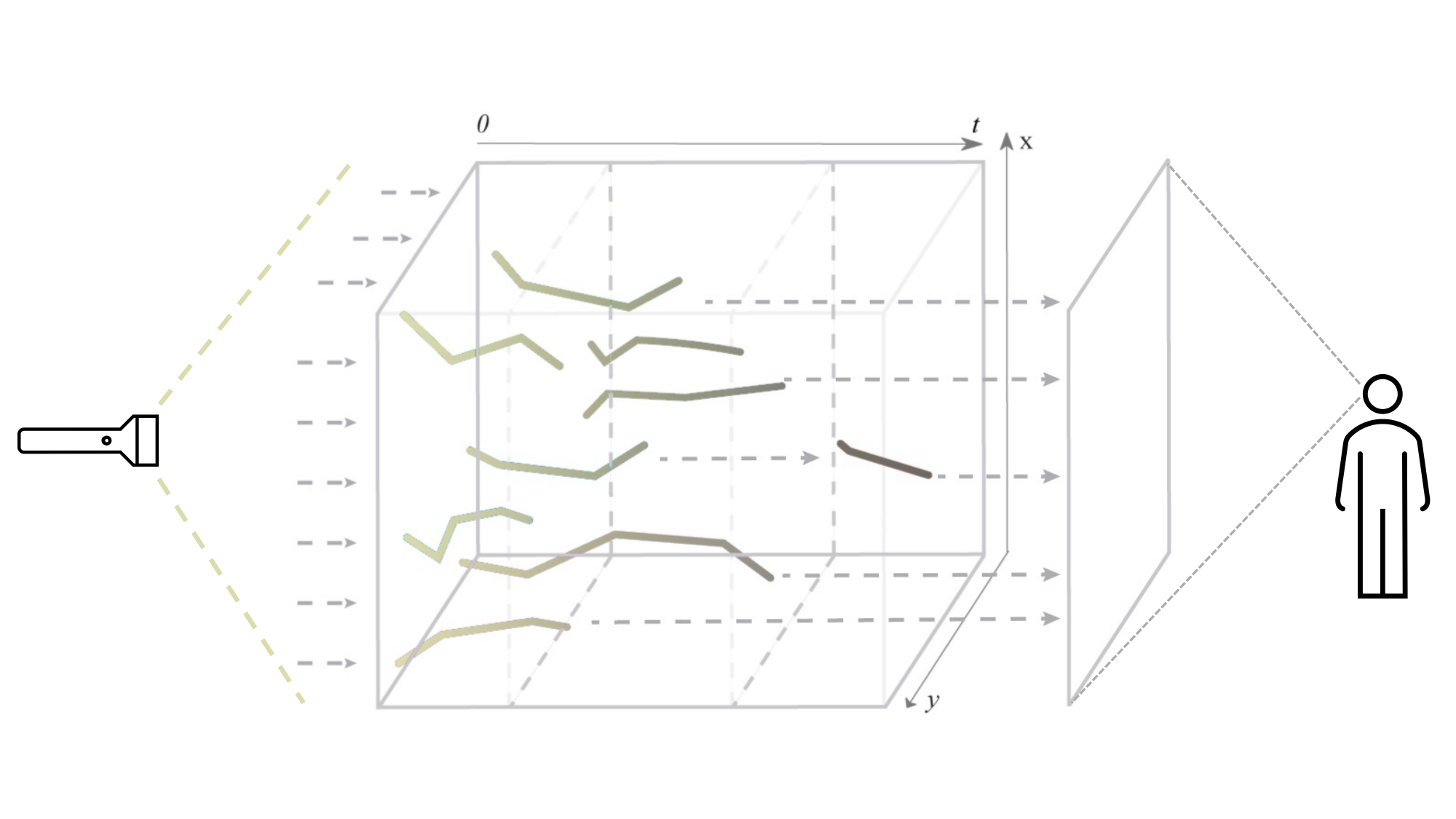}
    \caption{The \sys metaphor. Rays of light (dashed lines), travel from $t=0$ through the space-time cube and reach the observer at $t_{max}$. They interact with nodes' trajectories and carry this information to the projection plane.}
    \label{fig:3dtorch}
\end{figure}

The design of \sys is inspired by the ``time-coloring'' operation~\cite{17Bach}, %first described by Bach et al.~\cite{17Bach}, 
where time is mapped to color to visualize the evolution of nodes and edges through the space-time cube; and is loosely motivated by \textit{transfer functions} used in direct volume rendering~\cite{ljung2016state} to emphasize features of interest in the data.
Conceptually, in our approach, we ``shine'' light through the space-time cube down along its time axis (hence the system's name -- \sys) in a manner that resembles the behavior of transfer functions for volume rendering. As the light interacts with the node trajectories, they are visualized and colored differently according to the node's age (i.e., applying the time-coloring operation), generating a 2D visualization of the 3D embedding (see also Fig.~\ref{fig:3dtorch}). The resulting visualization is an explorable 2D map of the nodes' densities, with visible individual node movement and ``aging'' over time. We complement this visualization with several interactive controls to explore the data and introduce a movement score to rank the nodes based on the length of their trajectory. We designed \sys introducing multiple elements of visual guidance~\cite{ceneda2017characterizing} to enhance (and possibly ease) the network exploration process.
Finally, we describe two case studies demonstrating how our guidance-enhanced approach supports users in achieving the system design tasks. In summary, our key contributions are:
\begin{itemize}
    \item A novel visual analytics approach, \sys, to visualize and explore temporal graphs embedded in the space-time cube (see Section~\ref{sec:timelighting}).
    \item Two case studies presenting the utility of \sys to extract insights from temporal graphs and how the guidance-enhanced features support this (see Section~\ref{sec:case-studies}).
    \item A two-fold expert evaluation of our proposed approach and the supplementing guidance features provided (see Section~\ref{sec:evaluation}).
    % \item A reflection on the outcomes of the evaluation and discussing the potential and impact of our approach for temporal graph visualization (see Section~\ref{sec:discussion}).
\end{itemize}

\section{Related Work}\label{sec:related}
%Before providing the details of our guidance-enhanced approach to temporal network exploration, w
%In the following, we illustrate related literature on which we ground our research.

\textbf{Visualization of Dynamic Networks}. The typical definition of a dynamic network implies the notion of \textit{timeslicing}: the temporal dynamics are represented as a sequence of snapshots (the \textit{timeslices)}, each one representing the status of the graph over an interval of time, placed at equally spaced moments~\cite{beck2017}. Several algorithms have been proposed for drawing dynamic networks by using timeslices~\cite{brandes2011quantitative}. Branke~\cite{Branke2001} presents an overview of methods in dynamic graph drawing in the early days of the field. 
Beck et al.~\cite{beck2017} broadly characterize techniques %propose a survey about dynamic network visualization based on this definition where they broadly categorize the design space 
into timeline (or ``time-to-space'') and animation (or ``time-to-time'')~\cite{aigner2011}.
%, with the latter being the most widely employed approach. 
%Timeline approaches present scalability problems for longer time series in terms of the required space to represent it all. Superimposing multiple timeslices within the same view may mitigate this problem and facilitate an overview of the changes occurring to the network~\cite{gleicher2011}.
%In terms of visualization paradigms, timesliced networks have been extensively visualized using different approaches to encode the temporal dimension. 
% Each individual time point is called a timeslice, a snapshot that represents the state of the graph over a time interval.
% This simple yet powerful simplification is used as the basis of visualization~\cite{beck2017,graphdiaries}, for layout algorithm design~\cite{erten2003graphael,brandes2011quantitative}, and in user studies~\cite{archambault16,archambault2012,10FarrugiaIVJ,FilipovTVCG2023TVCG}.
Kerracher et al.~\cite{kerracher2014} introduce a two-dimensional design space of dynamic network visualization along two categories: graph structural and graph temporal encodings. %In their work, they outline that embedding temporal and structural information within the same view is a scarcely explored area. 
%Following this definition, 
Filipov et al.~\cite{filipov2021exploratory,filipov2022GD, FilipovTVCG2023TVCG} explore this design space by comparing adjacency matrices and node-link diagrams, with different temporal encodings (i.e., juxtaposition, superimposition, animation, and timeline). The qualitative study suggests the combination of animation and node-link representation as the better-performing option for the tasks tested.

Farrugia and Quigley~\cite{10FarrugiaIVJ} measure the performance of animation and timeline dynamic network visualization on node-link diagrams. They find that timeline representations were faster than animation and preferred by users.  Archambault et al.~\cite{archambault2011} found that timeline approaches were faster with no difference in error rate for most tasks, but some evidence for animation in helping with appearance tasks. 
Archambault and Purchase~\cite{archambault16} looked at the mental map's impact in low-stability settings (i.e., significant changes between subsequent timeslices) for node-link dynamic graph drawings by comparing animation and timeline. However, small multiples were consistently faster than animation with no significant difference in error rate for most tasks in this series of experiments.
%A mental map is defined as the mental model that the user creates while exploring the network and allows them to orient themselves during the exploration. 
%Their results highlight that in a low-stability setting, helps via animated transitions.  

%Extending to $2D + t$ or $3-$dimensional representations, 
For matrix representations, Bach et al. explore visualizing dynamic networks by stacking chronologically ordered adjacency matrices, effectively using the third dimension as a time axis~\cite{bach2014}. Windhager et al.~\cite{windhager2020} explore the use of space-time cubes to provide an overview of multifaceted data over time. %Our work focuses on node-link approaches.
%Furthermore, dynamic networks have been visualized in $2D + t$ using animated transitions to switch from a space-time cube to an animation over the series of timeslices, superimposing them within the same view, and juxtaposing them in a side-by-side manner.

Many approaches for online dynamic graph drawing have been proposed where a single timeslice is modified online with no knowledge of future changes~\cite{frishman2008,cohen1992,Ellson2004,gray2023}. However, we consider offline dynamic graph drawing whereby the algorithm has full access to the information across the entire time period of interest during drawing. 

%As we anticipated in the beginning, all these approaches model time as a discrete axis. None have considered investigating continuous (temporal) network data and the challenges it presents. 
All the above techniques use the timeslice as a basis for visualizing the dynamic network. Visualizing temporal networks %~\cite{12Holme} 
differs significantly from how typically we draw and visualize dynamic graphs in the graph drawing and visualization community as no such sequence of timeslices necessarily exists.%where the time axis is a discrete series of timeslices. 
%In the area of dynamic graph drawing and visualization, 
The problem with timeslicing is that many networks of scientific interest do not have natural timeslices, rather edges and nodes come in with their own specific time coordinates. Therefore, choosing a regular sampling and duration %for each can be a complex task, which eventually
often results in the loss of temporal information.
%Furthermore, selecting an appropriate amount of timeslices and a duration for each is also a non-trivial task that may obscure the dynamics of the network by aggregating a large number of events together.
In response, visualization techniques have been proposed for interesting timeslice selection. Wang et al.~\cite{NonuniformTimeslicing} present a technique for non-uniform time slicing (that is, selecting slices of different duration) based on histogram equalization to produce timeslices with the same number of events. % and, in turn, similar visual complexity. 
Lee et al.~\cite{Lee19Plaid,lee2020} experimented with methods for interactive time slicing on large touch displays.

Algorithms now exist that draw temporal networks directly in the space-time cube~\cite{simonetto2018event,arleo2022event}. These approaches do not impose timeslices on the temporal network but directly embed the network in the space-time cube. However, the only way to visualize such $2D + t$ drawings on a 2D plane was to select timeslices or present an animation of the data over time. The work we present here bridges this gap for temporal network visualization.

%Well-known limitations of animation are that they are not effective for some tasks involving dynamic graphs~\cite{archambault2011}, and exploring visualizations of these drawings is beneficial.  We draw inspiration to improve existing work by performing a ``time flattening'' and ``time-coloring'' operation~\cite{bachstc2014}, where we project the entire extent of the temporal graph and its dynamics onto a 2D surface to depict its dynamics. 
%Our proposed method could be viewed as {\it time-coloring}~\cite{17Bach} whereby time is mapped to color in order to visualize the evolution of nodes and edges in the graph through the $2D + t$ space-time cube drawing of the network data.  In some ways, it is similar to shining a flashlight down the time axis of the space-time cube, hence the name TimeLighting.

\textbf{Guidance for Graph Exploration}
Due to the challenge of analyzing complex events such as those modeled by temporal networks, researchers also investigated approaches to provide support and ease the analysis for users. The resulting approaches fall under the definition of ``guidance''~\cite{ceneda2017characterizing}. Guidance is characterized as \textit{active} support in response to a \textit{knowledge gap} which hinders the completion of an interactive visual data analysis session. Over the years, several approaches have been devised, providing different types of guidance~\cite{ceneda2019review,ceneda2022TVCG}. May et al.~\cite{may2012using} describe a method to enhance the exploration of large graphs using glyphs. While the user explores a given area of interest (the focus), the system automatically highlights the path to other possibly off-screen interesting nodes (the context). Gladisch et al.~\cite{gladisch2013navigation} provide support during the navigation through large hierarchical graphs by suggesting what to explore next. %Thanks to a user-customizable degree-of-interest function, the system can suggest how to navigate the graphs, both horizontally and vertically, adjusting the level of abstraction of the hierarchy. 
Despite the work in this area, applying guidance to temporal networks is uncharted territory. Given a temporal network drawn in a space-time cube, our goal is to provide guidance to support the identification of interesting time intervals and nodes requiring further attention and analysis. %In summary, existing literature reveals a gap in addressing continuous (temporal) network data challenges. Our \sys approach bridges this gap by offering a visualization method and guidance techniques.

%\section{Space-time Cube Time Coloring}
\section{Design Considerations}\label{sec:design}
In this section, we discuss the most relevant aspects that influenced the design of \sys, namely, the data characteristics, the users and tasks~\cite{miksch2014}, and the time-coloring paradigm used to provide guidance.

\subsection{Data, Users, Tasks.} 
%\noindent\textbf{Data, Tasks.}
%We design TimeLighting following the ``Data-Users-Tasks'' design triangle by Miksch and Aigner~\cite{miksch2014}. This approach bases a VA application design on three principles: the \textit{Data} the application will have to elaborate, the \textit{Users} who will interact with the system, the \textit{tasks} it will perform. We elaborate on these aspects in the following.
\noindent \textbf{Data}. \sys aims to support the visualization, analysis, and exploration of interval temporal network data~\cite{12Holme}. In a temporal network $D = (V,E)$, $V$ is the set of nodes, $E$ is the set of edges. Each node and edge comes with time-dependant (expressed as $t \in \Re$ in the following) \textit{attributes}. These take the form of functions in the $V \times t$ and $E \times t$ domains for nodes and edges, respectively. For simplicity, and in accordance with existing literature~\cite{simonetto2018event,arleo2022event}, we consider all attribute functions as piece-wise linear functions. The \textit{appearance} $A_x$ attribute, for example, models the intervals in time in which nodes and edges exist:

%The temporal network can be defined as $D = (V, E, T)$: a node $v \in V$ or edge $e \in E$ existing at a certain point in time $T$, depending on the \textit{appearance ($A$)} attribute as in the following:
\vspace{-0.9em}
\begin{align*}
A_v : V \times t \rightarrow [true, false]\\
A_e : E \times t \rightarrow [true, false]    
\end{align*}

$A_v$ and $A_e$ map to the node and edge insertion and deletion \textit{events}, respectively. For this reason, these graphs are also called \textit{event-based} networks, and the terms, temporal and event-based, will be used interchangeably in the remainder of this paper. The \textit{position} $P_v : V \times t \rightarrow \Re^2$ attribute describes the nodes' position over time. The following is an example of how each node's ($v \in V$) position is computed by an event-based layout algorithm describing the movement over time and space:

\begin{equation*}
    P_v(t) = \begin{cases}
            (5,6) \rightarrow (12,11) & \text{for } t \in [0,1] \\
            (8,3) \rightarrow (1,7) & \text{for } t \in [5,7] \\
            \cdots & \\
            (0,0) & \text{otherwise}
        \end{cases}
\end{equation*}

Therefore, each node draws a \textit{trajectory} in time and space. While a timesliced drawing would require computing the node coordinates at each point in time, in a temporal layout these trajectories must be computed and optimized in the three dimensions of the space-time cube. It is a complex and computationally intensive operation: in this paper, we use the \texttt{MultiDynNoS}~\cite{arleo2022event} event-based layout algorithm to generate the drawings of the temporal graphs.

\noindent \textbf{Users}. Temporal graph drawing is a difficult concept to comprehend and put into practice. With \sys we attempt to lower the bar of the necessary expertise to deal with these drawings by introducing guidance (see Section~\ref{sec:guidance}). However, in our system, we still target users with expertise in graph drawing. We, therefore, primarily aim to support the target audience of dynamic network visualization experts and support them in the analysis of temporal graphs. We further consider their feedback and discuss directions for future work on how to adapt the system to support novice users in extracting insights from temporal graphs.

%\textbf{Tasks.} 
\noindent \textbf{Tasks}. We derive \sys tasks from the task taxonomy for network evolution analysis by Ahn et al.~\cite{ahn2014} and the taxonomy of operations on the space-time cube by Bach et al.~\cite{17Bach}. The specific tasks are discussed in the following, along with a short description and justification.

%is designed for several user tasks, which we characterize using the task taxonomy for network evolution analysis by Ahn et al.~\cite{ahn2014} and the taxonomy of operations on the space-time cube by Bach et al.~\cite{17Bach}, and are described in the following.

\textbf{T1} - \textit{Overview.} The system should provide an overview of the temporal information at a glance. Providing an overview is typically the first necessary step in any VA process. This task considers the entire graph as an entity of the analysis~\cite{ahn2014} and is related to the orthogonal ``time'' flattening operation~\cite{17Bach}.

\textbf{T2} - \textit{Tracking Events.} Understanding the temporal dynamics and the events' frequency helps the user isolate interesting occurrences in time. Events also cause the nodes' trajectories to bend, i.e., make the node change direction. Understanding the \textit{shape}~\cite{ahn2014} of changes in nodes' movement over time (e.g., speed, repetition, etc.) would provide further insights during the exploration of the data. 

\textbf{T3} - \textit{Investigate relationships.} Each edge occurrence perturbs the trajectories. Identifying which relationships have the most impact or how often they occur might help explain the formation of clusters, or, in general, the phenomenon at hand. This task aligns with the ``point extraction'' task described in Bach et al.'s taxonomy~\cite{17Bach} since we are selecting an individual node and observing how it behaves over time (i.e., changing neighborhood over time).

% Any tasks explicitly related to the temporal dimension? (i.e., finding specific points / intervals in time?) -> "Time Chopping" by Bach et al., Temporal Features of individual Events by Ahn et al.?

%We also additionally calculate the trajectories of each node's movement over the entire temporal extent of the network, resulting in two distinct categories of nodes. The path length of the individual trajectories represents a mobility score that can be used to guide and highlight nodes that are extremely mobile or stationary over the course of the network's evolution. 

%The position and layout of of the network was calculated using \textit{MultiDyNoSlice}~\cite{arleo2022event} which was then exported into the popular \textit{GML} format and parsed into a \textit{JSON} file to be used with \textit{d3.js} for the development of the interactive visual interfaces (described in the following). 

%To address the challenges identified in Section~\ref{sec:related}, we introduce TimeLighting, a guidance-enhanced VA approach for the exploration of temporal networks. In the following, we illustrate the design choices that guided its development.

%\subsection{Time-coloring and TimeLighting Intuition} 
\subsection{A 2D Projection of a Temporal Graph}
To highlight interesting nodes and trajectories and more generally to visualize the network's dynamics, we took inspiration from the time-coloring operation described by Bach et al.~\cite{17Bach} in their survey. Time-coloring is a content transformation operation applied to the space-time cube. Given a timesliced graph, the procedure consists in coloring each timeslice based on a uniform linear color scale to facilitate the identification of the ``age'' of the data points.

To produce our 2D projection of a temporal graph, we took inspiration from the time coloring operation. However, to extend this technique to temporal graphs, we need to find a way to visually represent two features of each node: their \textit{movement} (i.e., its behavior during its appearance intervals) and their \textit{aging}. This means analyzing how their movements are distributed in time from the point of view of the observer, in a continuous time-axis scenario.
%Similarly to transfer functions for volume rendering, such 

To solve this problem, we propose the following solution. If we projected light through the cube along the time axis, the interaction between the light and the node trajectories will be visible from the observer, watching from the other side of the cube (at $t_{max}$, see Fig.~\ref{fig:3dtorch}). 
The edges of the trajectories closer to the light will appear brighter, with the side closer to the observer will appear darker, reflecting their aging process respectively (older--transparent; younger--opaque). Concerning the node's movement, we perform \textit{time flattening}~\cite{17Bach}. The results are 2D trajectories that ``narrate'' both the age and movement of the nodes over time. Inevitably, some temporal information is obfuscated in a 2D projection: however, we integrate it in the visualization through auxiliary views, interactions, and \textit{guidance}~\cite{ceneda2017characterizing} to support the users during exploration.
%In our scenario, the network needs to be projected, at first, in a 2D space (i.e., a \textit{time flattening} operation). However, we do so in a way that considers the polylines that represent the nodes and surfaces that represent the edges.
%Subsequently, its exploration can be enhanced with guidance suggestions, highlighting interesting time intervals (and nodes) to explore (a \textit{temporal coloring} operation). 
% To support these tasks, we visualize the network as a node-link diagram. The temporal dimension is visualized as a timeline.

\subsection{Guidance} \label{sec:guidance} 
In addition to visualizing the temporal network, we designed guidance to support its exploration and analysis. In general, the degree of support provided to the user may vary significantly, and at least three guidance degrees can be identified~\cite{ceneda2017characterizing}: 

\begin{itemize}
    \item \textit{Orienting} helps users keep an overview of the problem and alternative analytical paths they can choose to proceed; 
    \item \textit{Directing} guidance provides users with a set of options and orders them according to their importance for solving current tasks; 
    \item \textit{Prescribing} guidance (as the name suggests) prescribes a series of actions to take to conclude the task.
\end{itemize}

%guidance aims to uphold the orientation of the user by providing them with a set of \textit{options} (e.g., nodes, edges, parameters, etc.) from which to choose. In other words, orienting guidance helps users keep an overview of the problem and the alternative analytical paths they can choose to move forward. \textit{Directing} guidance provides users with a set of options and orders them according to their importance for solving current tasks, giving directions to steer the analysis. Finally, \textit{prescribing} guidance (as the name suggests) prescribes actions to take to conclude the task.
Given the similarities between our problem and the transfer functions used in volume rendering (although we apply them to a 2D visualization), the general idea at the base of our guidance-enhanced approach is to support and ease the identification of time intervals and nodes with specific desirable characteristics. The aim is to investigate their relationships and how they interact (i.e., analyzing the moment of the trajectories), making them stand out from the rest for the user's convenience. Considering our set of design tasks, we highlight the nodes that have the longest trajectories. Long trajectories represent nodes that are associated with many events and with high movement in the currently selected temporal interval (to support \textbf{T2}). This type of guidance can be classified as directing, as trajectories are ranked based on their length, and, is necessary to ease or solve the system design tasks.
In addition, we highlight temporal intervals in which nodes defined by the user interact with each other,
%interactions among a set of user-defined nodes happen, 
thus providing guidance to \textbf{T3}.

\section{\sys}\label{sec:timelighting}
\begin{figure*}[t!]
    \centering
    \includegraphics[width=\textwidth]{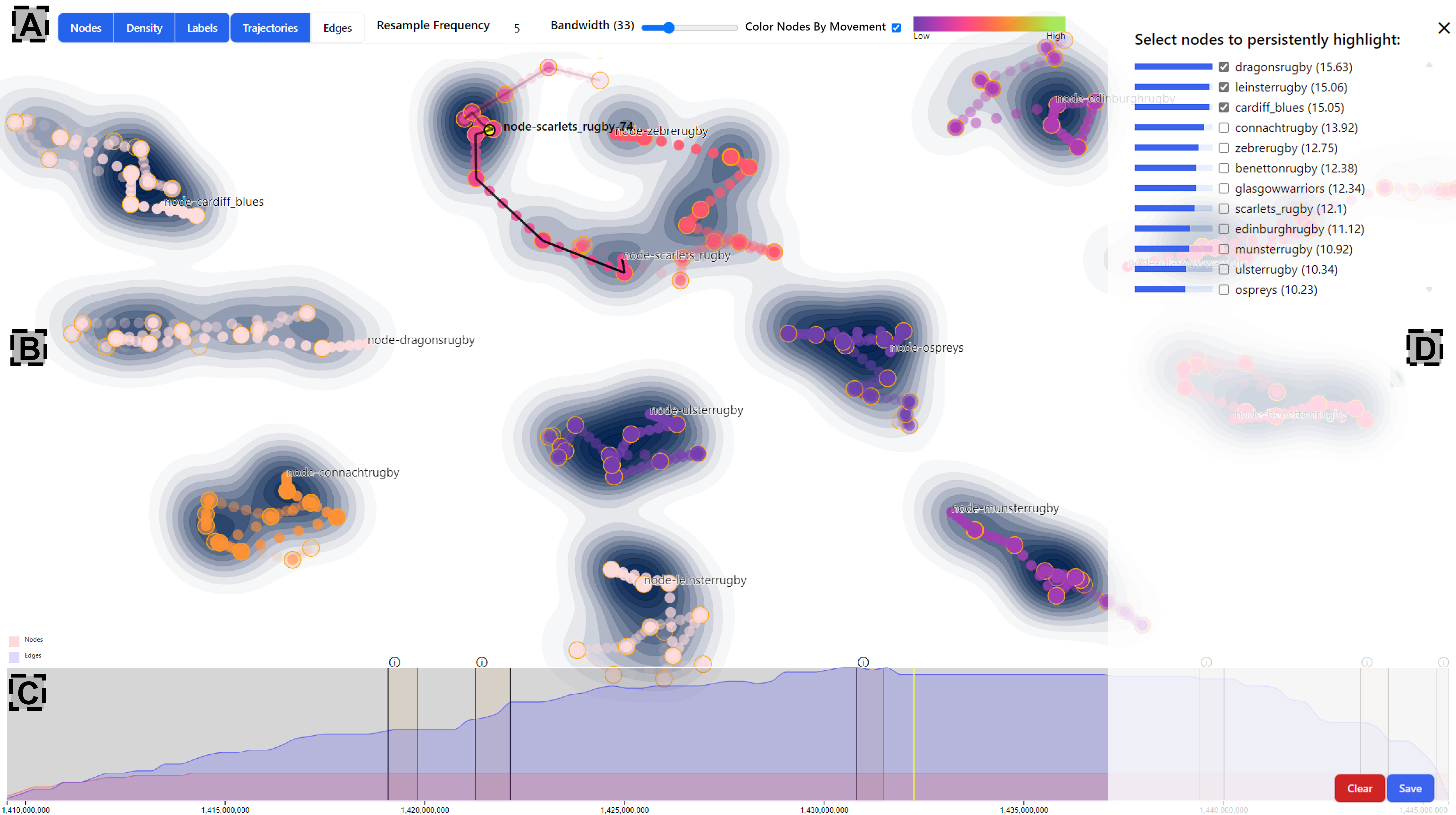}
    \caption{\sys overview. The view is comprised of the (A) toolbar and sidebar, (B) main view, (C) event timeline, and (D) the side bar depicting the movement scores. 
    %Trajectory sampling is set to 10 points. 
    The yellow bar in the timeline shows the absolute age of the hovered node (visible in the top-center area).}
    \label{fig:overview}
\end{figure*}

In this section, we describe our system in detail and how we implemented it considering the design requirements described in the previous section.
\sys is a guidance-enhanced VA system comprised of two linked views and a focus+context approach. An overview of the prototype is shown in Fig.~\ref{fig:overview}.  For more details and resources we refer to the online supplementary material~\cite{miro}. The source code is available on \href{https://github.com/velitchko/timelighting/}{\faGithub \xspace GitHub}. A demonstrational video showcasing the interactions is available online on \href{https://www.youtube.com/watch?v=GqBbqNR07rA}{\faYoutube YouTube}.

\subsection{Main View}

In the main view of \sys (see Fig.~\ref{fig:overview}-B), we show a 2D projection of the complete temporal graph (i.e., an overview---\textbf{T1}). We discuss the details of our approach in the following. We employ several different encodings to highlight features of nodes and edges.

\begin{figure}[t!]
    \centering
    \begin{subfigure}[b]{0.48\textwidth}
         \centering
         \caption{}
         \includegraphics[width=\textwidth]{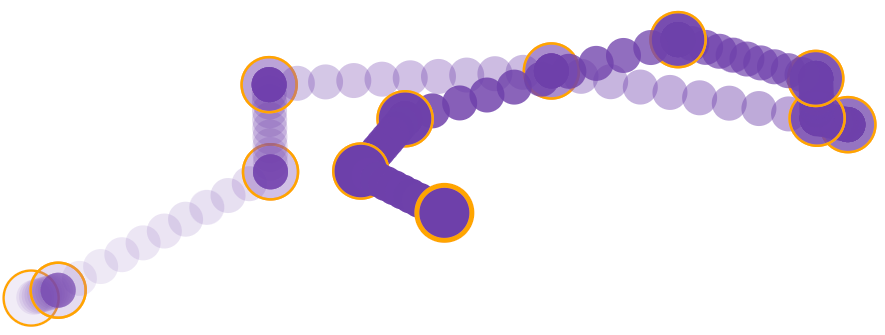}
    \end{subfigure}
    \hfill
    \begin{subfigure}[b]{0.48\textwidth}
         \centering
         \caption{}
         \includegraphics[width=\textwidth]{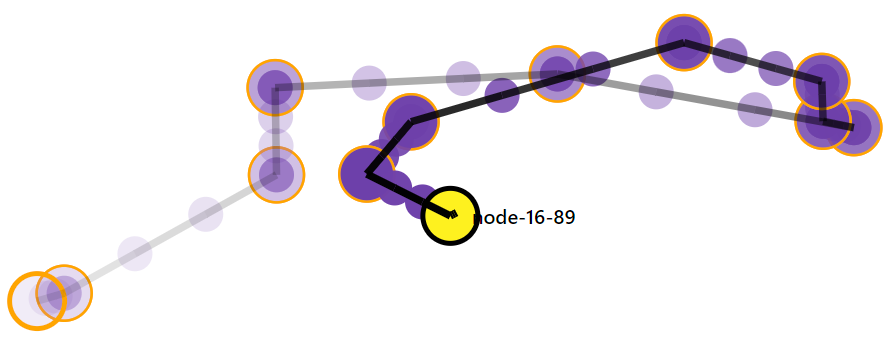}
     \end{subfigure}     
    \caption{Trajectories: (a) a higher sampling frequency is selected and aging is visible thanks to the change in opacity; (b) sampling is lowered (3 points per segment) and the trajectory is superimposed on mouseovering the yellow node. Orange stroke and size difference indicate coordinates coming from the data compared to interpolated nodes.}
    \label{fig:traj_comparison}
\end{figure}

\noindent \textbf{Node Positions} are represented as trajectories encoded as a trail of circles (see Fig.~\ref{fig:traj_comparison}). First, we place each interval's start and end position within every node's $P_v$ attribute (see Section~\ref{sec:design}). These come directly from the computed drawing and have an orange stroke to make them distinguishable from the sampled nodes. To ease the comprehension of the movement flow (\textbf{T2}), between the start and end positions of each interval, we place several sampled nodes, where the coordinates are interpolated. The user can fine-tune the number of interpolated positions by choosing an appropriate ``sampling frequency''.
%: in turn, the system samples the node movement a number of times equal to the sampling frequency.
%at regular intervals when a sampling frequency is selected. 
The resulting sampled nodes are positioned along the node trajectory but are encoded as smaller circles with no stroke to differentiate them from non-sampled nodes. 
We calculate and visualize the node \textit{aging} as follows: for each node visualized on screen (nodes both in $P_v$ and interpolated ones are considered), its age is computed as the difference between its time coordinate and the time of the node's first appearance. We use a linear opacity scale to visually represent the node's aging process.
%over the temporal extent of the network 
This encoding makes it easier to understand the information about the progression and movement of each node over time, providing an overview of the evolution of the network. We use relative aging in this context as it is focused on the individual node's trajectory. Hovering over the node makes it possible to see its position in the timeline (in the context of the full temporal extent of the network) as a yellow bar, corresponding to its time coordinate (see Fig.~\ref{fig:overview}-C).
Typically, nodes are visualized in gray. However, users can change the visual appearance of the nodes to reflect the cumulative amount of movement (see Section~\ref{sec:guidance}--\textit{Guidance}). Activating this type of guidance changes the coloring from gray-scale to a continuous color scale making nodes with higher movement visually distinct from the more stationary ones (\textbf{T2/T3}).

\noindent \textbf{Edges} are represented as solid straight lines that connect pairs of nodes belonging to two distinct trajectories. Edges have a pivotal role because these interactions eventually cause node movements and the creation of temporal clusters (\textbf{T3}). Edges might also appear or disappear within $P_v$ movement intervals. This justifies the choice of introducing sampled nodes: edges can appear between all the points belonging to a trajectory, including interpolated ones. This allows the user to keep an overview of the finer temporal details, as we can display edges closest to their exact time coordinates.
%, as edges might appear or disappear within $P_v$ movement intervals.
In turn, this could generate visual clutter as each edge is shown once for every node pair between each trajectory, and, depending on the sampling frequency, a trajectory can be comprised of several nodes. We mitigate this by showing edges on-demand and related to one trajectory at a time similar to an ego-centric approach (the selected one). The user, by hovering on any node belonging to a trajectory, will make all edges incident to those nodes appear. Edge aging is encoded similarly to node aging. Fig.~\ref{fig:case_study_2}-B illustrates an example of how edges connect the sampled nodes within the current temporal selection for the currently hovered node (Munster Rugby). 
%that are within the selected interval in the timeline.

%Their appearance $A_e$ (see Section~\ref{se:definitions}) is also
%the age of each edge and encode this in a linear opacity scale similar to the node aging in order to maintain a consistent visual representation of the aging. The edge's age is calculated in a similar fashion to the nodes, in that it is the difference between the current point in time where the edge exists and it's first appearance.

\noindent \textbf{Movement} is visualized using a polyline connecting each position in the nodes' $P_v$ attribute. It represents how (and how much) the node's position changes over time due to the bends in the trajectories computed by the layout process. To clarify, the \textbf{movement} score of a node is calculated as the total distance traveled by the node over the time period for which it is present, this the sum of the Euclidean distances between consecutive positions of the node's $P_v$ attribute. This score directly correlates with the trajectory length; hence, longer trajectories indicate higher movement scores.

This movement (trajectory) encoding supplements the nodes' opacity as opacity alone might not be sufficient in showing how trajectories evolve (\textbf{T2/T3}). We calculate the age of each trajectory segment and apply a linear opacity scale (as with the nodes and edges). The trajectory's age is calculated as the mean of the ages of each pair of nodes in that given polyline segment. This information is also shown on-demand by hovering over a node, and, either the edges or the movement of a node can be seen (top bar selection---see Fig.~\ref{fig:overview}-A). 
%Our approach is inspired by Dimpvis~\cite{kondo2014dimpvis}.

\noindent \textbf{Density} is represented as a contour map (see the dark-blue areas in the center of Fig.~\ref{fig:overview}), providing a quick visual indicator that emphasizes locations where a larger number of nodes have existed (\textbf{T2}). This kind of encoding also provides a first glance at the trajectories' ``shape'', which eases keeping an overview of the events (\textbf{T1}). To calculate the density map, we translate the original set of nodes for each point in time into a series of objects with $x, y$ coordinates and relative age. The $x, y$ coordinates determine the contours of the density map. The age acts as a weighing function such that older nodes would contribute less to the density map compared to more recent nodes. The ``bandwidth'' sets the standard deviation of the Gaussian kernel, with lower values showing a sharper picture and higher values more distributed, but also more blurred, representation.
%is interactively configurable and updates the density map accordingly.

% \noindent \textbf{Interactions} %\noindent \textbf{mouseover Highlighting:} To interact with the individual nodes in the main view, we have mouseover highlighting that updates the color of the currently selected node to red, as well as, shows either the trajectory or edges to adjacent nodes depending on the current state of the controls. 
%\noindent \textbf{Zoom \& Pan:}
% are also supported in the main view of TimeLighting. Common interactions such as using the mouse scroll wheel to \textit{zoom} and rescale the main view, and \textit{panning} or \textit{dragging} to reposition it. 
% This interaction is intended to support further investigation on specific areas of the network. %(zoom and details on demand). 

%\noindent \textbf{Sidebar Guidance:} In the side bar as previously described we sort all nodes by their mobility score in a descending manner and enable the top 3. This default selection is intended to be a type of guidance to point to the more interesting nodes in a data set (i.e., the high mobility nodes). The supplementing bar chart representations for each node in the list is also designed to support this idea. In cases where a larger selection of nodes (more than 3) is of more interest the bar charts can be used to identify which nodes have a higher mobility score. 

\subsection{Other Views and Guidance}

\noindent \textbf{Top Bar.} At the top of the prototype (see Fig.~\ref{fig:overview}-A), a set of controls allows users to configure parameters and the appearance of the visualization. All individual encodings can be toggled on or off (nodes, labels, and density map). 
% There are also controls to change the bandwidth of the contour map and a checkbox to select the gray or movement-dependent coloring of the nodes.

Furthermore, depending on the specific analysis or exploration task, the nodes' trajectories or edges between them can be toggled as visible on mouseover. Additionally, we have a configurable re-sampling frequency that will recompute the nodes and their coordinates and update the visualization accordingly when changed. A modifiable kernel size allows the user to fine-tune the the contour density map. Finally, there is a checkbox that controls the state of the node coloring. By default, the nodes are colored gray and their opacity represents their age. By toggling the check box the coloring is updated to reflect their movement over time using a continuous color scale, with opacity still representing their age.   

%\noindent \textbf{Main View.} The main view represents the entire network (see Fig.~\ref{fig:overview}-B). In this view we have the nodes being layed out according to the coordinates from the \textit{MultiDyNoSlice}~\cite{arleo2022event} algorithm and the edges and trajectories can be seen on mouseover interactions on each of the nodes (described in more detail in the interactions section). Furthermore, under the nodes, edges, and trajectories is where the density map is drawn. In this way it is possible to match the position and age of a set of nodes with a more denser hot spot in the contour density map.

\begin{figure}
    \centering
    \includegraphics[width=\linewidth]{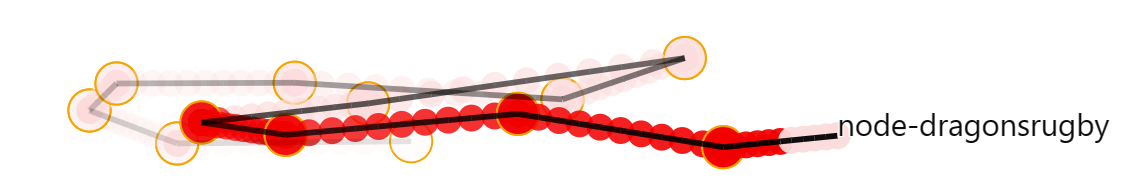}
    \caption{Example of a locked trajectory. The circles in red are fully within the temporal selection of the users, whereas the less saturated ones are outside of the selected interval.}
    \label{fig:red_nodes}
\end{figure}

\noindent \textbf{Side Panel} shows a list of nodes ordered according to their movement scores (see Fig.~\ref{fig:overview}-D). This ordering acts as guidance (see Section~\ref{sec:design}) by highlighting nodes that had many interactions with other nodes and that could, therefore, be considered interesting. Each node in this list is also accompanied by a small bar chart visualizing the differences in the movement scores. From this panel, the user can select and ``lock'' trajectories in the main view.
%in a visual manner.
%, and their trajectories are ``locked'' in the main view. The user can refine or change this initial selection.
%As additional guidance, when loading a new graph, the top 3 nodes are locked by default
A locked trajectory is always shown regardless of the current temporal interval selection in the timeline described in the following. The purpose of locked trajectories is to support users with a selection mechanism to focus on and examine specific node behaviors in detail over time, even as other aspects of the network and visualization change. This is particularly useful for tracking the movement and interactions of the locked (selected) trajectories without losing their context in the overall network. 
For example, if a researcher is analyzing a social network to understand the behavior of highly mobile individuals, they might lock the trajectories of these key nodes. This allows them to observe how these individuals interact with others over time, how relationships are formed and dissolved, and how their movement influences the rest of the network. By locking these trajectories, the researcher can maintain a continuous overview of these nodes' activities, providing insights into patterns and anomalies that might be missed if the trajectories were not consistently visible.
% In this case, when locking a trajectory we can observe how it starts to interact with others in the network (as relationships are formed and dissolved) while preserving the temporal overview of the locked trajectory.
%regardless of their temporal behavior. 
Locked nodes are colored in bright red if they are in the current temporal selection, while nodes out of the current temporal selection are colored in a less saturated hue (see Fig.~\ref{fig:red_nodes}).
%differently to help the user discern the and those out of it. 
%If a selected node is within the current temporal selection it is colored bright red, whereas if it is outside of this selection, it is colored in a lighter red. 
The encoding and ordering of node trajectories in the side panel serve as visual guidance to support the tracking of specific events (i.e., guidance to ease \textbf{T2}). Additionally, when loading a new graph, the three nodes with the highest movement are locked by default (this can later be refined or changed).

\noindent \textbf{Timeline}, shown in (Fig.~\ref{fig:overview}-C), allows users to select and explore specific temporal intervals as well as keep an overview (\textbf{T1}) of the number of nodes and edges that are visible over time. This overview is obtained by considering the number of nodes and edges for which the appearance attribute ($A_v$ and $A_e$) is \textit{true}. We visually encoding this information as two overlapping area charts showing the changes occurring to the network as it evolves (in red for the nodes and blue for the edges).

The timeline serves two purposes: first, the user can brush to select a specific interval within the available data. As a result, only the subgraph existing during the newly selected time interval will be shown in the main view (temporal filtering). This also affects movement coloring, relative age, and density calculation, as well as, limits the edges shown on-screen to those existing in the current selection (these do not apply to locked trajectories). Second, \sys uses the timeline to provide guidance and suggest specific time intervals for further inspection. Specifically, the system highlights intervals in time when all the currently locked trajectories interact with each other. These intervals are represented as orange rectangles drawn on top of the timeline (see Fig.~\ref{fig:overview}-C). Clicking one of these intervals will snap onto that temporal selection, helping the user to keep track of and investigate relationships (i.e., guidance to ease \textbf{T3}).

%The brush can be dragged in order to select these intervals for more detailed inspection. Additionally, we also provide some guidance to intervals of time that may be more interesting for in-depth exploration based on the currently selected nodes in the side panel. We superimpose these intervals as orange rectangles in the background with an icon that snaps to that current window. The intervals are identified as start and end points in time where all of the selected nodes exist or appear simultaneously. The intent of this guidance approach is to facilitate a more intuitive and straightforward approach to selecting intervals of time that are potentially of interest in the analysis and exploration process.
%(ii) it is used as a filter out to select specific intervals of time that should be investigated in more detail. Upon selecting a specific range in the timeline the main view is updated to only depict nodes, trajectories, and edges that appear in this range. The density map is also updated accordingly to reflect new hot spots within this selected interval.

\subsection{Interactions}
\noindent \textbf{Highlighting:} The individual nodes in the main view can be interacted with by mouseover, which highlights the currently selected node (yellow) as well as showing either the trajectory or edges to adjacent nodes depending on the current configuration. Furthermore, when mouseovering a specific node its current location in time will be highlighted in the timeline using a yellow bar (see Fig.~\ref{fig:overview}-C).

\noindent \textbf{Zoom \& Pan:} The main view also supports common interactions such as using the mouse scroll wheel to zoom and rescale the main view and panning or dragging the main view to reposition it. This interaction is intended to be used when only subsets of the network in the main view should be investigated in more detail. 

\noindent \textbf{Sidebar Guidance:} In the sidebar, we sort all nodes by their movement score in a descending manner and lock the top three. This default selection is intended to be a type of guidance to point to the more interesting nodes in a data set (i.e., nodes with high movement). The supplementing bar chart representations for each node in the list are also designed to support this idea. In cases where a larger selection of nodes (more than three) is of more interest, the bar charts can be used to suggest which nodes should be selected (i.e., higher movement scores).  

\noindent \textbf{Timeline Guidance:} The timeline can be used to identify intervals where a large number of changes to the number of nodes or edges are occurring. These are considered very dynamic intervals in the evolution of the network. The area charts showing the number of edge and node events over the entire period of time serve as a visual indicator to identify interesting intervals (i.e., high event frequency). Furthermore, the orange rectangles (see Fig.~\ref{fig:overview}-C) that are superimposed on the timeline guide specific intervals where all of the locked trajectories exist at the same time, supporting a more detailed analysis of the current selection. 

\section{Case Studies} \label{sec:case-studies}
In this section, we discuss two case studies on two real temporal networks. We show how insights can be extracted from the data using \sys and how the design tasks are achieved and supported with guidance. We build our case studies on the \textit{Rugby}~\cite{pro12} and \textit{Reality Mining}~\cite{realitymining} datasets and are described in the following.

\subsection{Case Study 1: \textit{Rugby} Dataset}\label{sec:cs1}

The \textit{Rugby} dataset is a collection of 3151 tweets posted during the Pro12 rugby competition of the 2014-2015 season~\cite{pro12}, specifically from September 2014 to October 2015. The network has a node for each team participating in the competition (12 teams in total), and an edge exists between two teams when a tweet from one mentions another. While nodes will stay visible from the moment they appear until the end, edges appear at the exact moment the tweet was posted. To improve the visibility of the edges during the layout process (as tweets do not have a ``duration''), edges are given a 24-hour duration. For example, if an edge $\bar{e}$ has a timestamp $t$, then $A_{\bar{e}} = [t - 12h, t + 12h]$. Multiple edges between the same teams are merged if their appearances overlap by a duration of less than one day. This simplification has already been applied in previous work using this dataset~\cite{simonetto2018event}, and discretization of this dataset with similar resolution would require 417 timeslices. This dataset is particularly interesting as we have ground truth data to validate our findings.

\begin{figure*}[!t]
    \centering
    \includegraphics[width=0.49\linewidth]{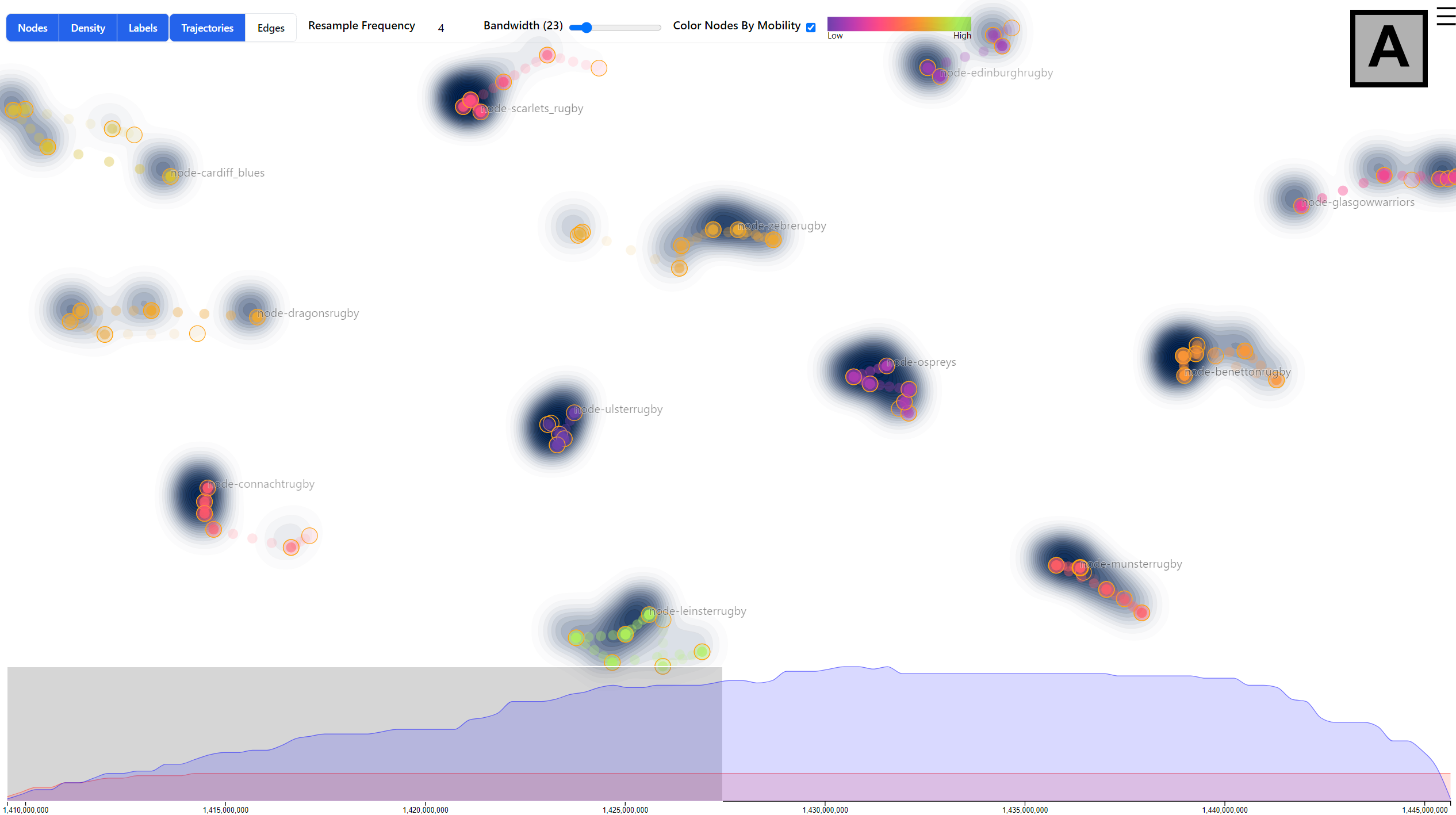}
    \hfill    \includegraphics[width=0.49\linewidth]{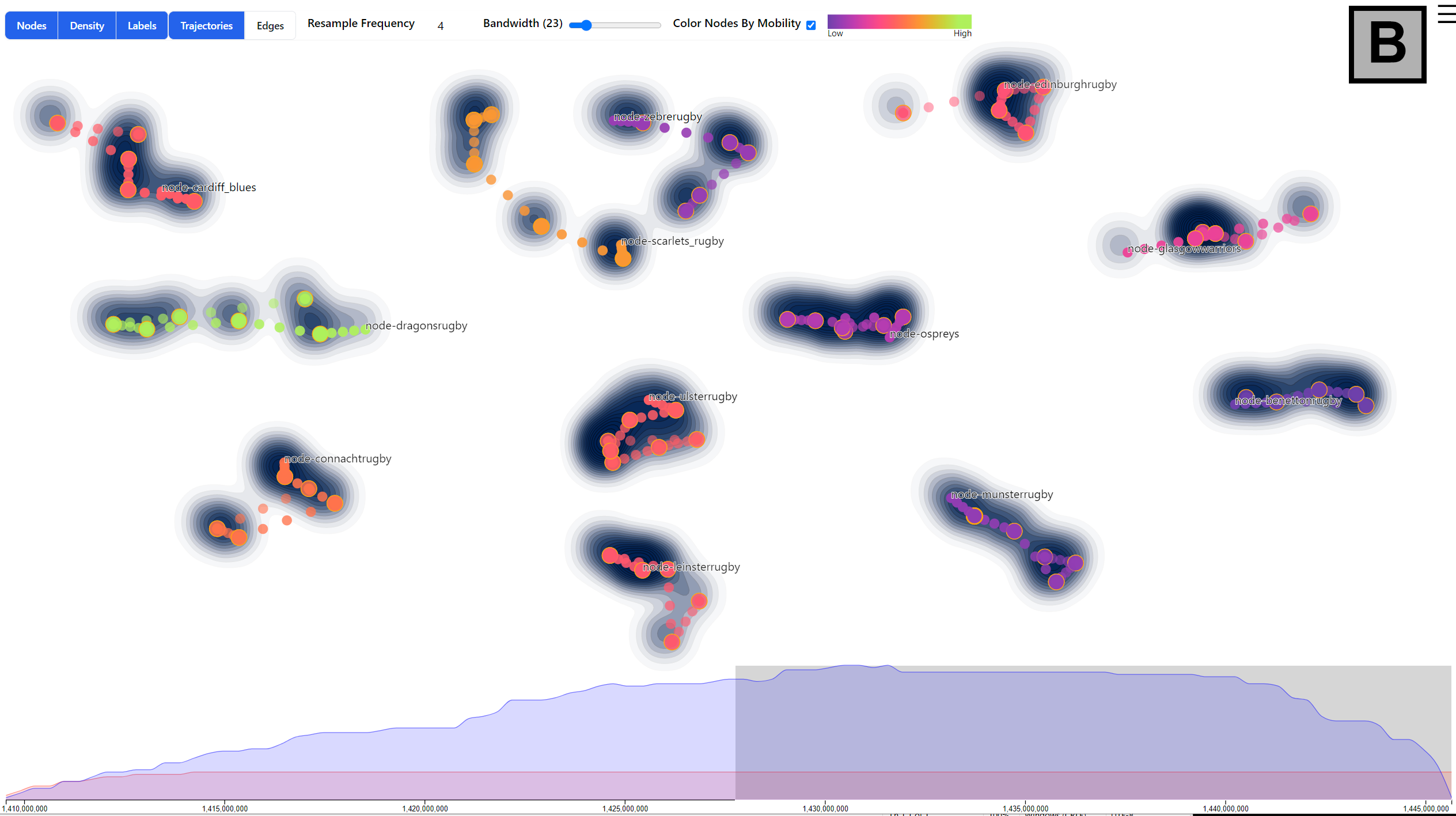}\\   \includegraphics[width=0.49\linewidth]{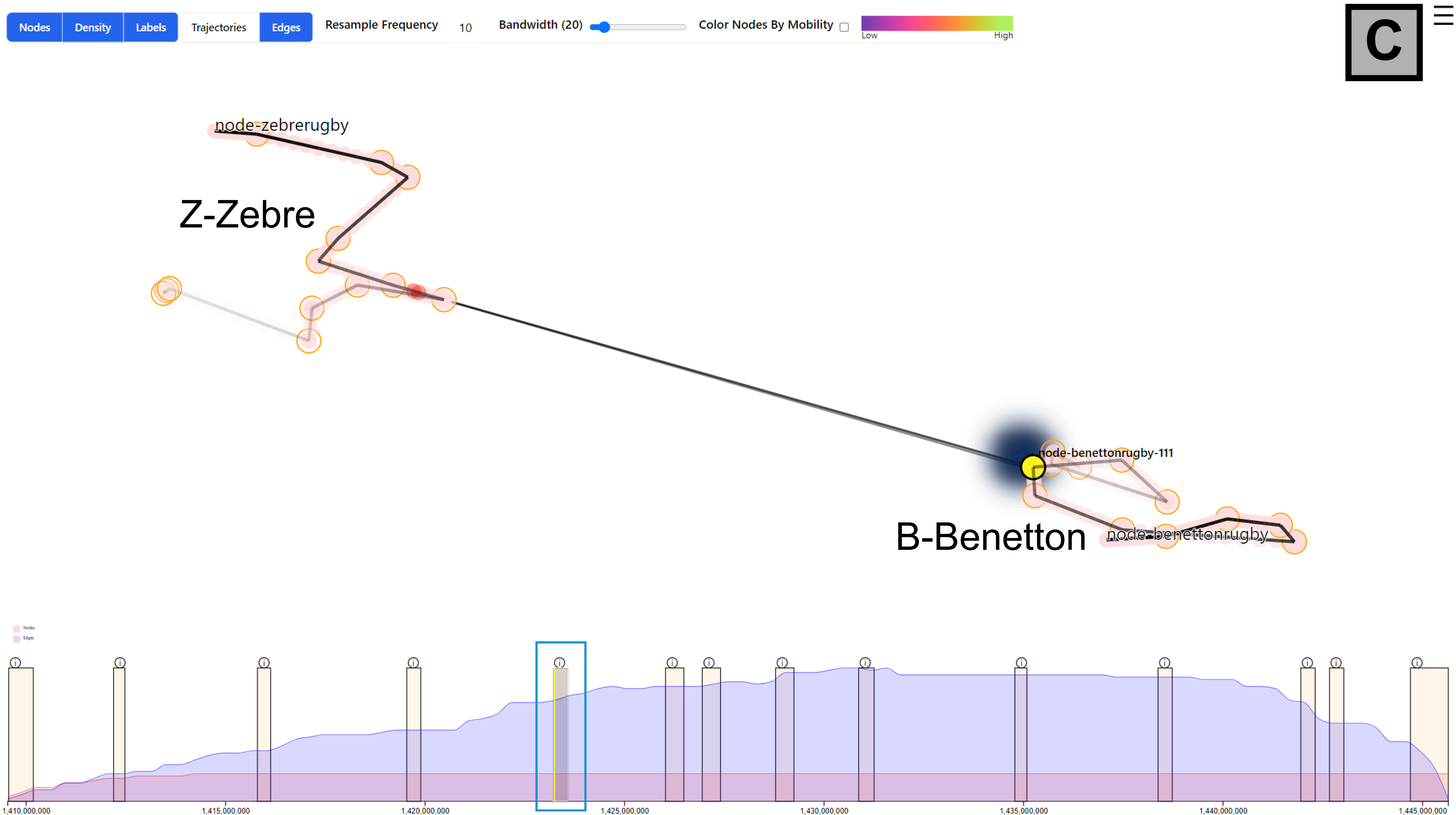}
    \hfill
    \includegraphics[width=0.49\linewidth]{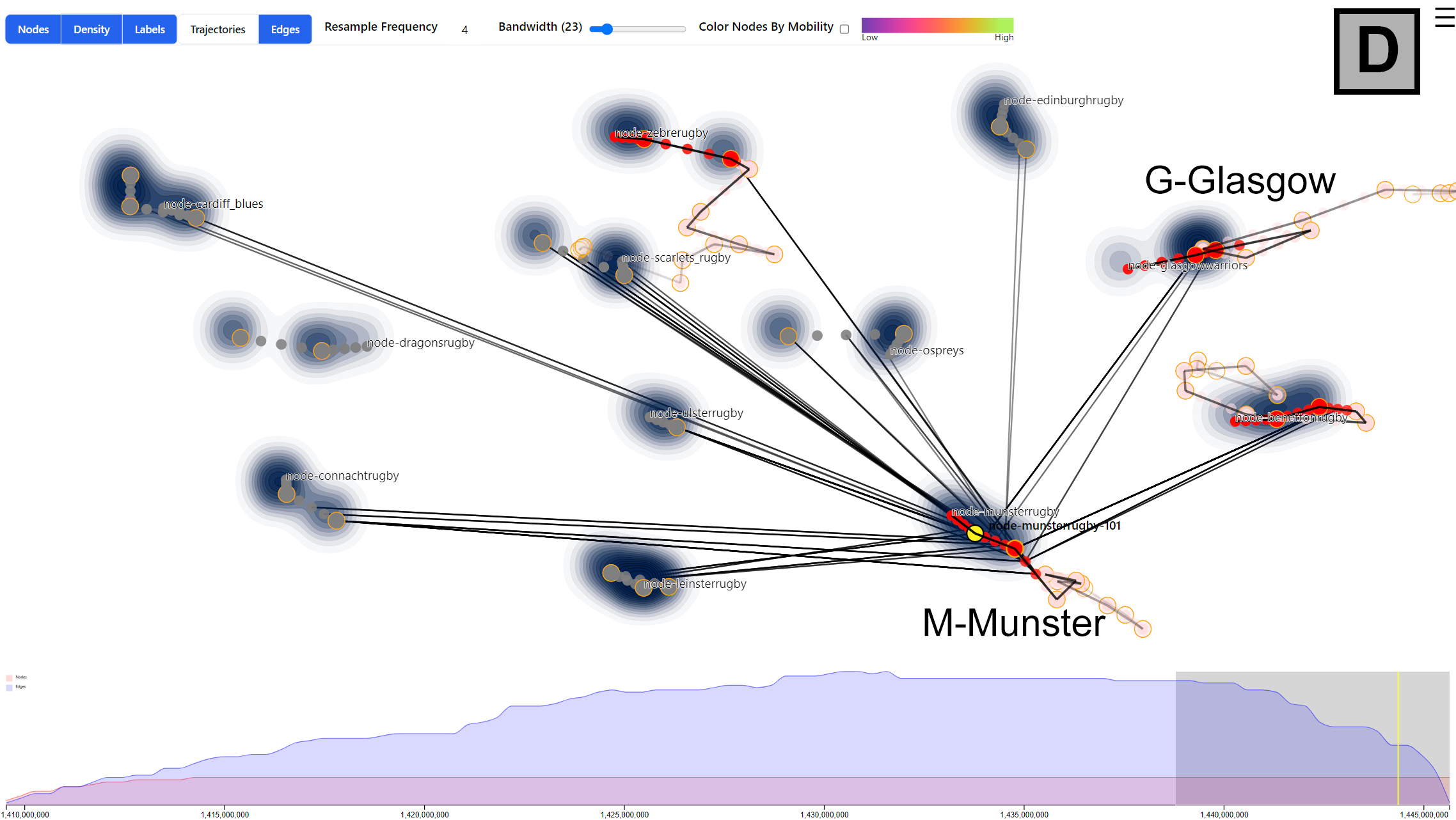}    
    \caption{Illustrations from Case Study 1. We can observe how the movement of nodes changes in the two halves of the season. \textbf{(A)} The first half of the season; \textbf{(B)} The second half of the season. Timeline brushing is used to filter out events. Trajectory sampling is set at 4 points. \textbf{(C)} The two teams' trajectories are visible as guidance intervals in the timeline. The selected one (blue rectangle) relates to the match between the two teams in the first round of the competition. \textbf{(D)} Highlights the final match and the connections between the 2nd best team and other teams of interest in the case study.}
    %\label{fig:case_study_1}
%\end{figure}
%\begin{figure}[!ht]
%\centering
%    \includegraphics[width=\linewidth]{figures/casestudy2.png}
%    \includegraphics[width=\linewidth]{figures/casestudy2_B.png}
%    \caption{Case Study 2: \textbf{(A)} The two teams' trajectories are visible as guidance intervals in the timeline. The selected one (blue rectangle) relates to the match between the two teams in the first round of the competition. \textbf{(B)} Highlights the final match and the connections between the 2nd best team and other teams of interest in the case study.}
    \label{fig:case_study_1}    
    %\label{fig:case_study_2}
\end{figure*}

\textit{The First and Second Half of the Season: Examining Trajectories.} 
We begin our use case with an \textit{Overview} task (\textbf{T1}), examining the trajectories in the first (see Fig.~\ref{fig:case_study_1}-A) and the second half of the 2014-2015 season (see Fig.~\ref{fig:case_study_1}-B). We can immediately observe from the timeline the trend of the events. There is a steady increase in the number of tweets from the beginning of the season that peaks around the beginning of the second round of the league. This peak remains until the season's final and significantly impacts the nodes' movement.
We can also see that nodes move less in the second half of the season compared to the first half. Specifically, in Fig.~\ref{fig:case_study_1}-B the majority of the nodes are within the purple to orange range of the scale---lower movement---whereas in Fig.~\ref{fig:case_study_1}-A they are in the yellow to green range---encoding higher movement. In the first half, instead, tweets are sparser, meaning that the influence of an edge on the movement of nodes is kept (as there is no inertia) until another one changes its trajectory. Continuing the analysis of the network, tweets happen at a much higher rate in the second half of the season, and, since this network is a clique (all teams eventually play against one another), they tend to be ``locked'' in place by the attractive forces exerted by the other nodes. 
It must be considered that the layout algorithm attempts to optimize (and reduce) node movement, placing the nodes in an area of the plane where they will likely remain. This behavior can be seen in the density map too, where hot spots are larger and more numerous (i.e., nodes tend to linger more in the same areas) in the second half compared to the first half. Nonetheless, the amount of attractive force will depend on the public interest (i.e., the number of tweets) about individual matches.

%\subsection{Case Study 2} 
\textit{Tracking the Two Least Winning Teams.}
In this second case study, we \textit{track} (\textbf{T2}) the trajectories and \textit{investigate} (\textbf{T3}) the relationships between the two least winning teams of the season (according to the historical information available), namely ``Zebre'' (Z) and ``Benetton'' (B). We begin by selecting them in the sidebar so that the whole trajectory is locked permanently on the screen. Guidance shows us the different moments in time when the two teams interact, and we focus on the period when the two teams play against each other around the midpoint of the season. The teams played two matches (during the first and second leg of the tournament) in adjacent rounds (12 and 13). The status of the interface is reported in Fig.~\ref{fig:case_study_2}-C. It is possible to see how the Z trajectory bends significantly towards B at this point, and a similar effect is visible the other way around. This attraction strength can be interpreted as the ``hype'' of the matches building up, as Z and B are the only two Italian teams in the competition. Finally, we compare the relationships between the last two teams in the ranking and the first two, ``Glasgow Warriors'' (G), the winners, and ``Munster Rugby'' (M), referring to the time around the tournament finals (see Fig.~\ref{fig:case_study_2}-D). If we focus on M, it is easy to identify the time the final was played, with mostly all teams connected to it. The B trajectory is strongly influenced by M, as it was one of the final matches before the final; Z, instead, is not largely influenced and drifts away.

\subsection{Case Study 2: \textit{Reality Mining} Dataset}
\label{sec:cs2}
In this second case study, we focus on the \textit{Reality Mining} dataset. The data comes from a study~\cite{realitymining}, where the mobile phones of the study participants were equipped with logging software. The study took place between September 2004 and June 2005, with the participants being students and staff from a major university. The complete dataset includes a wide variety of events, such as short messages, voice calls, local Bluetooth connections, and association with cell towers. For our case study, we only consider voice calls and text messages. Therefore, our nodes are the identifiers of the individual contacts (as they are referred to in the original dataset), and each edge between them is either a call or a short message. We use the real duration of the calls and set a fixed duration for the short text messages. Moreover, node appearance and presence are taken from the original data. For readability, we limit the number of nodes to 170, which results in a total of 333 events (edges). We modified the \texttt{MultiDynNoS} algorithm parameters to increase the movement of nodes. This network is more complex than \textit{Rugby}, and therefore we take advantage of the guidance and filtering tools of \sys to build our story. 

\begin{figure*}[!hht]
    \centering
    \includegraphics[width=0.49\linewidth]{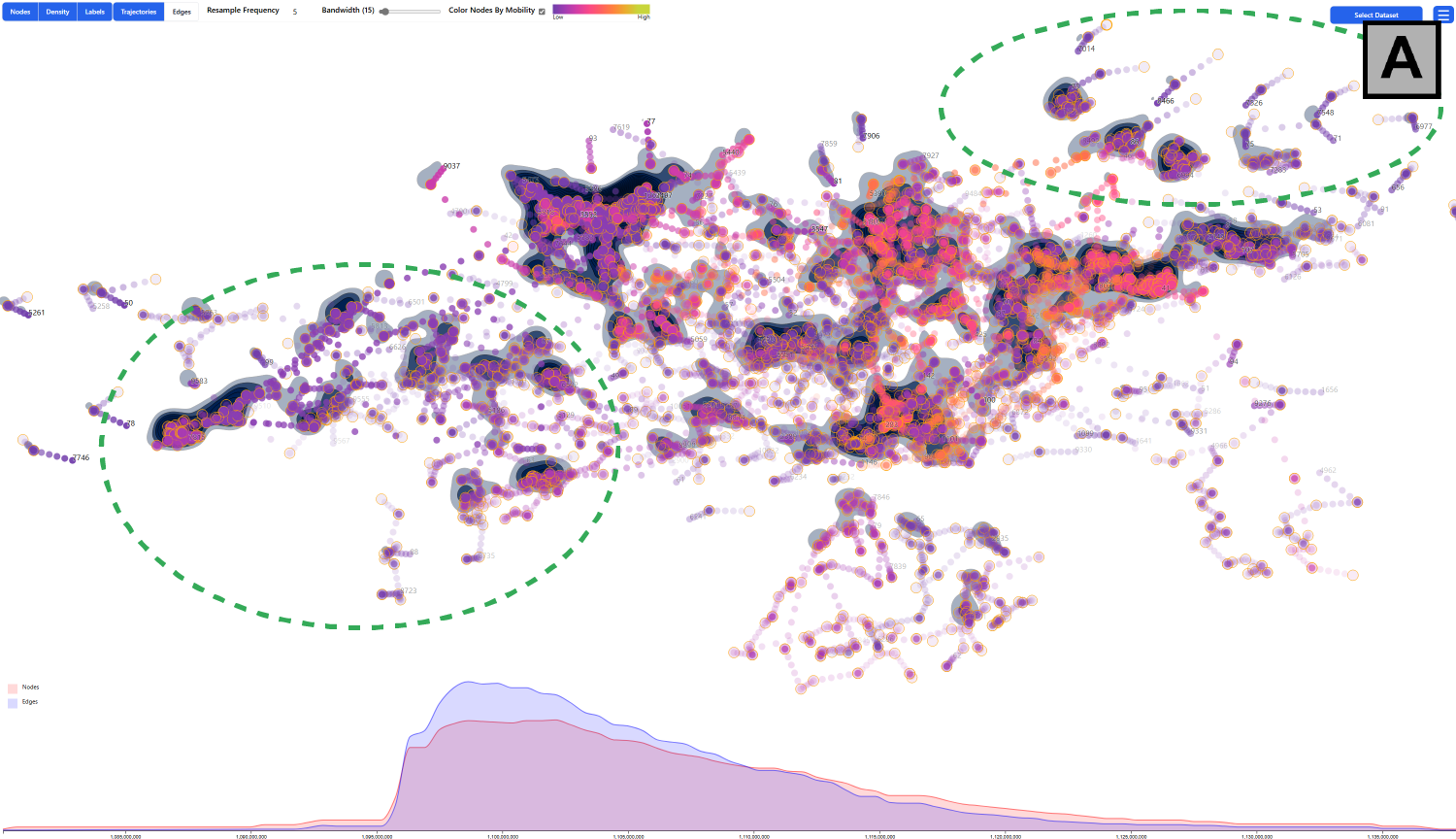}
    \hfill    \includegraphics[width=0.49\linewidth]{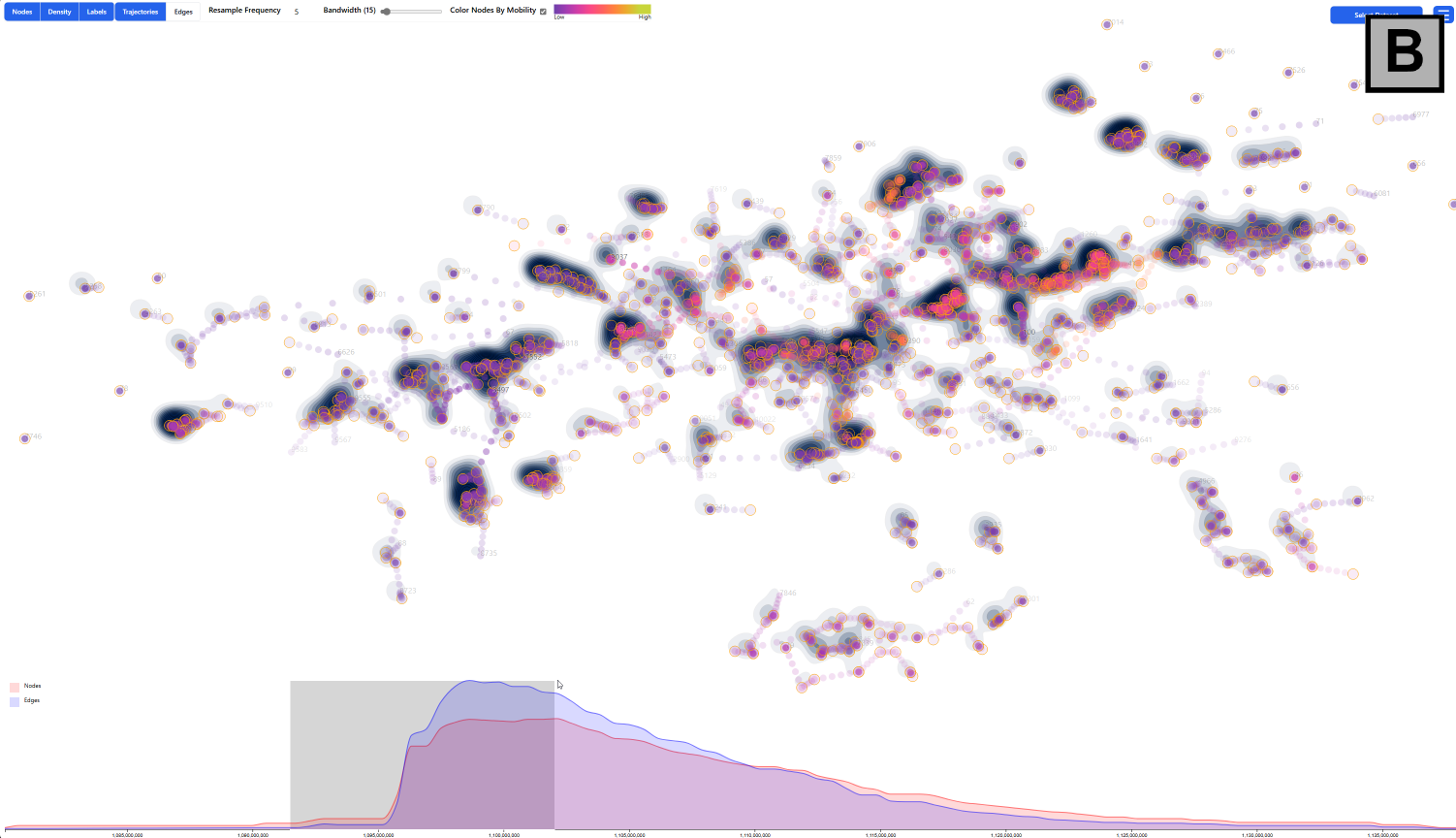}\\   \includegraphics[width=0.49\linewidth]{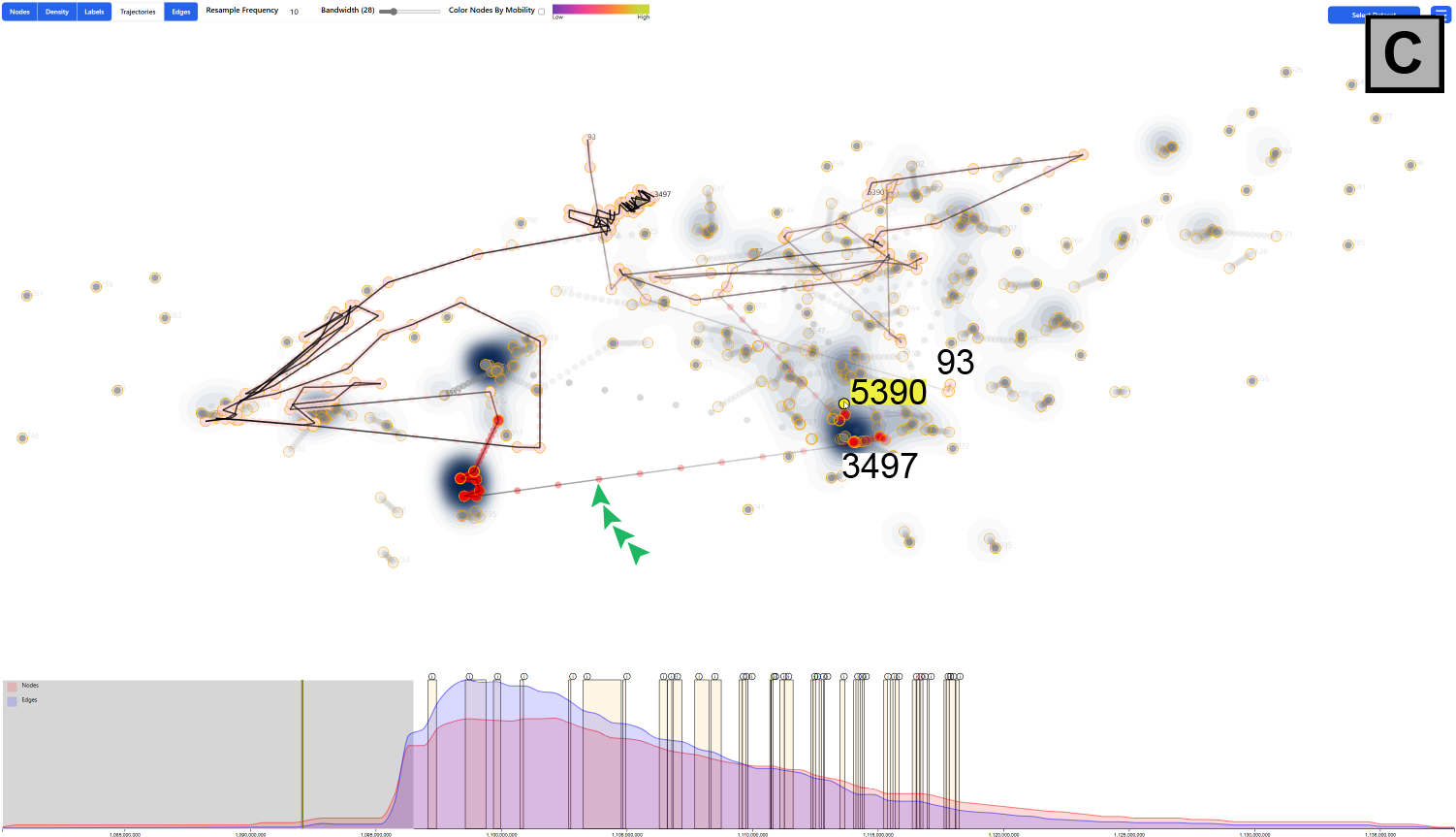}
    \hfill
    \includegraphics[width=0.49\linewidth]{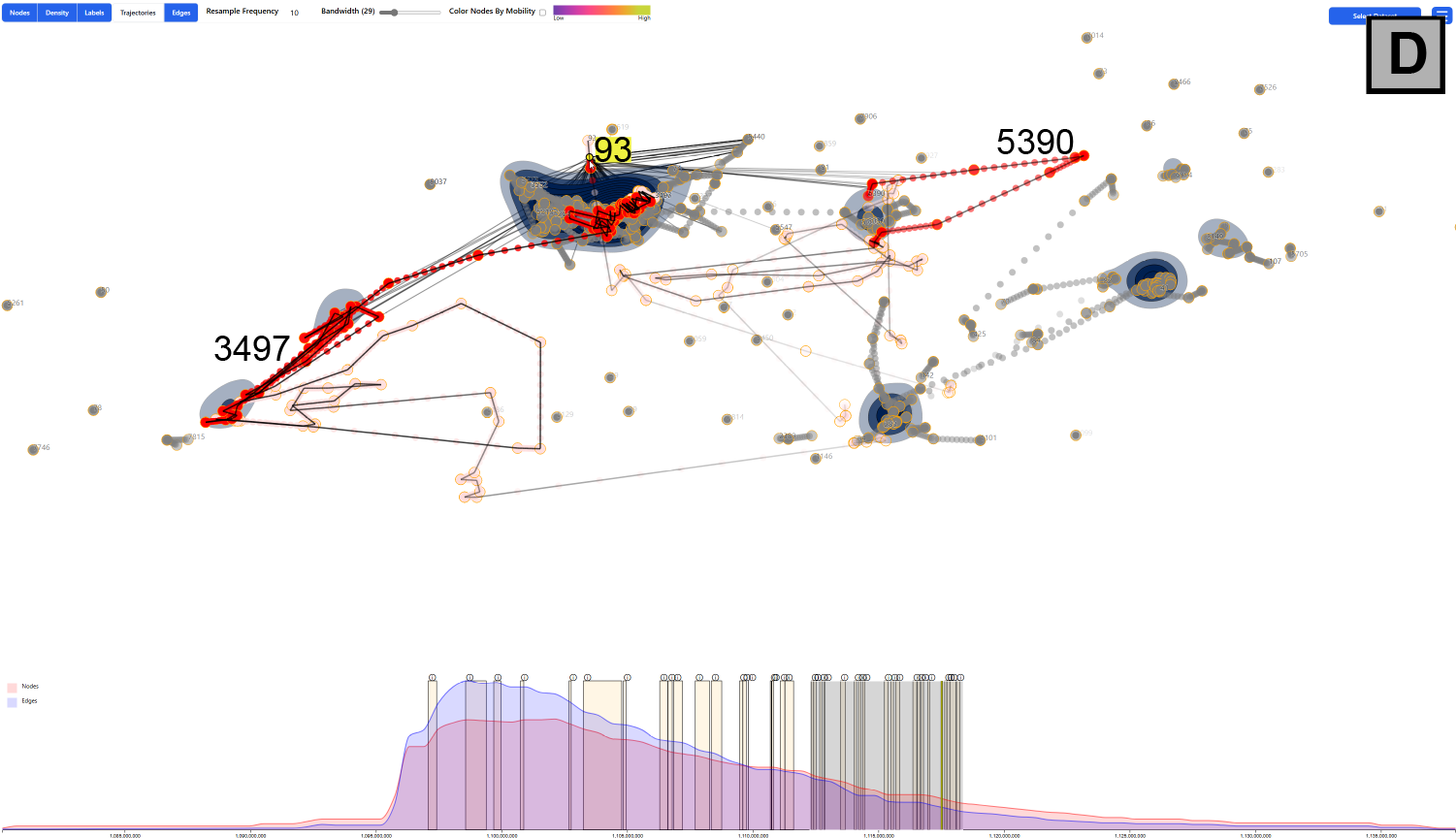}    
    \caption{Illustrations from Case Study 2. The complete projection of the dataset is shown in (\textbf{A}), with the two sub-communities referenced in Section~\ref{sec:cs2} have been highlighted with green ellipses. In (\textbf{B}) the cores of these communities are shown as they begin to form. In (\textbf{C}) the nodes trajectories have been labeled for readability. The green arrow indicates the position of node 3497 as 5390 appears. Finally, in (\textbf{D}) we highlight the positions of nodes 5390 and 3497.}
    \label{fig:case_study_2}    
\end{figure*}

\textit{Investigating communities.} The status of the system can be seen in Fig.~\ref{fig:case_study_2}-A. 
We set the resampling frequency to 5 (i.e., the number of sampled nodes between the start and end positions, see Section~\ref{sec:design}) to reduce the clutter and the kernel size for the density to 15 to improve cluster separation. Nodes are colored according to their movement. From the timeline, it is possible to spot that the majority of traffic happens around the first third of the timeline, it peaks and then subsides. We argue that it is the beginning of group projects or courses, as sub-communities form and a few entities participate in more than group one across the entire timeline, as we will see in the following (\textbf{T1}). It is easy to spot the different sub-communities: while some are more tightly connected, i.e., the large group in the center, the group on the left and top right belong perhaps to smaller courses or friendship circles (see Fig.~\ref{fig:case_study_2}-A). If we filter out the moments around the peak, we can see the cores previously mentioned communities forming (see Fig.~\ref{fig:case_study_2}-B). Moving the selected time interval forward, we see ``beyond'' the overview, meaning that as the semester progresses students come and go (i.e., appear and disappear from the dataset) and that groups reshuffle. Nodes move about significantly: the rate of movement can be inferred by the relative distance between the resampled nodes. A larger distance means more movement, as it must be covered in the same time interval indicated in the timeline.

\textit{Elements changing groups.} We now track the two nodes with the greater movement (node 5390---highest and node 3947---third highest). The node with the second largest movement followed a similar trajectory to 5390, and we could not make meaningful observations. We, therefore, focus on 3947 and select the portion of the timeline from the beginning to the peak. Our first finding is that, at the very beginning, 5390 and 3497 interacted with the same nodes from one group, but not between each other directly. Node 3497 is the first to appear and shortly after departs from the group. Node 5390 appears later, and remains in the group until the peak before leaving as in Fig.~\ref{fig:case_study_2}-C (\textbf{T2}, \textbf{T3}). Following the proposed guidance, it is possible to find other instances where this behavior repeats. If we focus on the area in the timeline indicating the peaks where multiple guidance intervals are located (see Fig.~\ref{fig:case_study_2}-D), we see that the two nodes interact with node 93. This occurrence repeats close to the ``tail'' of the dataset: around this time, node 3497 is interacting with the nodes remaining in the cluster, and interacts again with node 93 and, coincidentally, with 5390. Afterward, these last two will permanently disappear, followed shortly after by 3490. Further investigation shows that 93 was the last node to appear between the three and followed a direct trajectory similar to 5390 with a similar duration. We argue that these nodes (5390 and 3497) were possibly tutors, as they visited several groups during the semester, with 93 being either a common friend or colleague. This example further shows how the system features and guidance are useful in untangling more complex temporal networks. 

\section{Evaluation} \label{sec:evaluation}
We evaluate our system through two heuristics: the ICE-T methodology by Wall et al.~\cite{wall2019} to assess the \textit{value} of our proposed visualization, and the guidance evaluation heuristic by Ceneda et al.~\cite{ceneda2023}.
%Our evaluation follows a heuristic approach to assess both the appropriateness and efficacy of the presented visual encoding as well as the quality of insights that could be extracted by our proposed guidance features.This ensures a
This methodology aims to provide a well-rounded appraisal of the design decisions regarding both the visual encodings and the guidance features,
%the entire system and how it 
thus also suggesting how well they support \sys design tasks (as in Section~\ref{sec:design}).
%that we present in Section~\ref{sec:tasks}. 
In the following, we 
%describe the evaluation methodology we employ, 
outline the setting and procedures we follow, and present our results. For further reading about the evaluation and results, we refer to our online supplementary material~\cite{miro}. 

\subsection{Methodology}
%When designing our evaluation methodology, we wanted to assess the efficacy of design of both our visualization and of the guidance features. 

%We employ a two-fold heuristic evaluation methodology aimed at assessing both the capabilities of our approach to support insight-driven exploration and generation, as well as, identifying the suitability of the proposed guidance features in supplementing the analysis process.
Our evaluation is based on two concepts, which we aim to assess: the \textit{value of visualization} and \textit{guidance quality}.

\noindent\textbf{Value of Visualization.} The concept of \textit{value} of visualization was introduced by Stasko et al.~\cite{stasko2014} as the ability to convey a ``\textit{true understanding}'' of the data, beyond what usability task-based tests are typically able to uncover. Temporal networks are interesting to the visualization community as they carry more nuanced information over timeslicing. Therefore, we aim to visualize this extra knowledge appropriately and expressively.
%To evaluate the value of our proposed approach
For this reason, we employ the ICE-T heuristic evaluation methodology proposed by Wall et al.~\cite{wall2019} - as is an evaluation framework that stems from Stasko's principles. The objective is to determine how appropriate and effective the visual representation and interactions are in responding to data-driven questions, generating insights, and inspiring confidence in the analysis results. The framework is divided in four visualization \textit{components}: \textit{Insight}, \textit{Confidence}, \textit{Essence} , and \textit{Time}. Each component is divided into a set of guidelines, which are again divided into a set of heuristics, formulated as actionable and rateable statements.

\noindent\textbf{Guidance Quality.} Ceneda et al.~\cite{ceneda2023} in their work investigate what makes guidance \textit{effective}. They identify 8 guidance quality criteria: each criterion is then divided into a set of heuristics, formulated as rateable insights. The paper proposes a different set of heuristics if they have to be rated by visualization users or guidance experts: depending on the evaluation perspective and purpose. In our evaluation, we use the user heuristics.

%To assess the proposed guidance features, we apply a practical evaluation methodology proposed by Ceneda et al.~\cite{ceneda2023}. The evaluation methodology focuses on quantifying the effectiveness of the guidance from a user perspective by evaluating different qualities that contribute to the overall usefulness of the guidance approach. 

Both the evaluation methodologies described above aim to evaluate, in a qualitative manner, different aspects of a visualization approach. Moreover, they follow a similar protocol, each of the study participants fill out a questionnaire evaluating each heuristic on a 1-7 Likert scale.
%using Likert scales. 
Hence they can be effectively combined to provide an assessment of the \sys overall value. We refer the interested reader to the respective papers for further reading~\cite{wall2019,ceneda2023}.

\subsection{Setting and Procedure}
We recruited 6 experts who voluntarily agreed to participate with no compensation. All participants had a strong background in dynamic network visualization as the intended audience of our proposed approach requires such expertise. Each participant was individually interviewed.

The interview sessions were structured as open-ended exploratory analysis scenarios lasting 1 hour. Each began with a 15-minute onboarding, detailing the visual encodings, interaction techniques, and guidance features of \sys. In the remainder, participants were asked to freely interact with the system on three different graphs of increasing topological and temporal density. We prepared a set of optional tasks (see supplemental material~\cite{miro}), mainly intended to bootstrap the exploration (as they were not required by the evaluation methodologies we use). The interviews were conducted online using a remote video conferencing application and the participants were asked to share their screen contents as they interacted with the prototype. We kept a record of each interview, detailing how the participants interacted with our approach. We also recorded the insights they gained, comments, and feedback they had about the visual encodings, interaction techniques, or guidance features. Participant voices and on-screen interactions were recorded with their consent. 

At the end of each interview, we asked the participants to fill out a questionnaire structured as a combination of the ICE-T survey~\cite{wall2019} and guidance evaluation methodology~\cite{ceneda2023}. The questionnaire was provided to the participants as an online form to complete at their convenience. 

%Our evaluation is structured as expert interviews. , with each participant using the system using a think-aloud protocol~\cite{lewis82}. We aim at both gaining in-depth insights into the questions and answers that the participants had while engaging with our proposed approach. We recruited participants with a background in dynamic network visualization as the intended audience of our proposed approach requires such expertise. A total of 6 experts were recruited to participate in our evaluation.

% \noindent\textbf{Data}
We prepared three datasets for the evaluation, of increasing complexity. All three datasets have been previously used in visualization evaluations - with Infovis being particularly familiar to our intended audience and study participants. 
% \begin{itemize}

    % \item
    \noindent\textbf{Rugby}: is a temporal network derived from tweets exchanged between January 2014 and 2015 about the ``Guinness Pro12'' rugby competition~\cite{simonetto2018event, pro12}. The teams represent the nodes and the relationships are derived from tweets - it is the same dataset as in case study 1 (see Section~\ref{sec:cs1}). \\
    % \item 
    \noindent\textbf{Reality Mining}: It is the same dataset used in Case Study 2 (see Section~\ref{sec:cs2}), albeit in a simplified version. Specifically, compared to the case study, we only include the first 200 events.\\
    % \item 
    \noindent\textbf{InfoVis}: is a co-authorship network for papers published in the InfoVis conference from 2008 to 2018~\cite{CiteVis_data}. Authors of a paper are connected in a clique at the time of publication. %Differently from the other two deatasets, this is divided into 10 timeslices (one per year).
% \end{itemize}

% \input{tables/components_icet}

% \begin{figure*}[!ht]
%     \centering
%     \includegraphics[width=.9\linewidth]{figures/components.png}
%     \caption{The evaluation results of the ICE-T components~\cite{wall2019} and the guidance~\cite{ceneda2023}. The scores are calculated as the average of the guidelines related to each component using a Likert scale (0-7). The dashed line indicates a score of 5, which corresponds to the minimum score for an appropriate and effective visualization.}
%     \label{fig:components}
% \end{figure*}
% Add online information and materials as links here 

\subsection{Results}

%In this section, we present the results of our evaluation, structured as expert interviews. We present the results of the questionnaires, highlighting the value of our proposed VA approach as well as assessing the utility and efficacy of the guidance features provided. Furthermore, we present the most notable comments and feedback that the participants discussed during the evaluation sessions. 

%\subsection{Questionnaires}
\begin{figure}
    \centering
    \includegraphics[width=\linewidth]{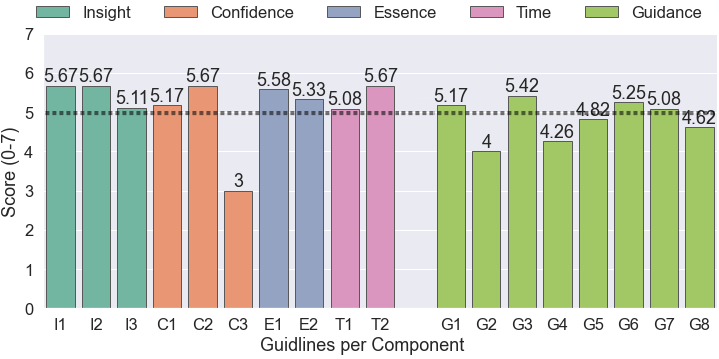}
    \caption{
    The evaluation results of the individual guidelines from the ICE-T~\cite{wall2019} and guidance evaluation methodologies~\cite{ceneda2023}. The scores are calculated as the average of the heuristics related to each guideline using a Likert scale (0-7). The dashed line indicates a score of 5, which corresponds to the minimum score for an appropriate and effective visualization. The complete definitions of the shortcodes for the guidelines are available in our online supplementary material~\cite{miro}.}
    % \textcolor{red}{While I like the idea, I believe the alignment provides a false sense of correlation - and I would include the per guideline scoring in suppl. material only}}
    \label{fig:guidelines}
\end{figure}

\noindent\textbf{Value of Visualization.} As described by Wall et al.~\cite{wall2019}, we averaged the questionnaire results per component. The component scores are in the range of 4.61 to 5.48 and the individual guidelines scores range from 3 to 5.67 (see Section~\ref{fig:guidelines}). A score of 5 is considered the minimum for a successful evaluation with the ICE-T method~\cite{wall2019}.

%The evaluation we conducted with network visualization experts for ``TimeLighting'' highlights notable strengths across several of the guidelines proposed by Wall et al.~\cite{wall2019} and Ceneda et al.~\cite{ceneda2023}.The responses and scores per guideline for each component are depicted in Fig.~\ref{fig:guidelines}, whereas the scores for each component are summarized and presented in Fig.~\ref{fig:components}. In the following, we present the results of the evaluation accompanied by a discussion of how the participants rated each guideline and component, weighing the strengths and weaknesses of ``TimeLighting'' according to the guidelines. 

% \input{tables/components_icet}
% \input{tables/components_guidance}
% \input{tables/guidelines_icet}

\textit{Insight component (Score \textbf{5.48}):} The participants rated \sys highly in terms of its ability to facilitate insight generation and provide new perspectives on temporal networks, scoring 5.67 (\textbf{I1}--\textit{The visualization facilitates answering questions about the data}) and 5.67 (\textbf{I2}--\textit{The visualization provides a new or better understanding of the data}), respectively. Additionally, the participants rated our approach with a score of 5.10 (\textbf{I3}--\textit{The visualization provides opportunities for serendipitous discoveries}) for providing opportunities for serendipitous discoveries, indicating its efficacy in revealing unexpected patterns within the data. Overall, \sys succeeds in effectively facilitating exploration and gaining an in-depth understanding of temporal network dynamics. Our proposed approach and the supporting interactions provide opportunities for insightful discoveries and gaining a better understanding of the underlying network patterns and trends.

\textit{Confidence component (Score \textbf{4.61}):} %Participants appreciated the visualization's capacity to help avoid making incorrect inferences
Overall, confidence guidelines were rated lower than the others. Participants rated
5.17 the visualization's capacity to help avoid making incorrect inferences (\textbf{C1}--\textit{The visualization helps avoid making incorrect inferences}), but they believe it facilitates broader learning about the domain with a score of 5.67 (\textbf{C2}--\textit{The visualization facilitates learning more broadly about the domain of the data}). Finally, they expressed a lower score of 3 (\textbf{C3}--\textit{The visualization helps understand data quality}) regarding its ability to understand data quality. This is somewhat contrary to the insights and observations made during the evaluation sessions, where the participants utilized the movement scores and mouseover interactions to identify unexpected behavior of the network.
%(related to data quality).
These scores show that this is a potential area for improvement to enhance how \sys could better support an immediate understanding of data quality and uncertainty in temporal networked data. 
%While our proposed approach supports users in avoiding incorrect inferences and broadens the understanding of the data domain, conveying data quality metrics is an issue that we can improve on.

\textit{Essence component (Score \textbf{5.46}):} \sys received a score of 5.58 for its capacity to provide a big-picture of the data  (\textbf{E1}--\textit{The visualization provides a big-picture perspective of the data}) and was ranked 5.33 on the ability to provide an understanding beyond individual data cases  (\textbf{E2}--\textit{The visualization provides an understanding of the data beyond individual data cases}), indicating its effectiveness in conveying overarching trends and patterns present in the data.

\textit{Time component (Score \textbf{5.38}):} Participants rated the visualization favorably in terms of its capabilities in offering rapid comprehension of the data with 5.08 (\textbf{T1}--\textit{The visualization affords rapid parallel comprehension for efficient browsing}) as well as enabling to quickly lookup and seek for specific information (data points) with 5.67 (\textbf{T2}--\textit{The visualization provides mechanisms for quickly seeking specific information}). The time component was rated highly by all the participants and this further highlights \sys efficiency in supporting users' exploration of temporal data.

%Finally, we also asked the participants to provide their thoughts and rate the guidelines associated with the guidance aspects of our approach.
% \noindent\textbf{Guidance}
\textit{Guidance component (Score \textbf{4.83})}: We averaged the score per guidance quality, with all of them scoring between 4.00 and 5.42--suggesting a positive final assessment~\cite{ceneda2023}.
The scores of the individual quality metrics highlight the strengths of our proposed features.
The proposed guidance was perceived as flexible as it adapted to the selection (i.e., selecting specific periods/windows of time or specific nodes to investigate in more detail) of the user (5.17, \textbf{G1}--\textit{Flexible: the guidance degree is adequate and adapts to the situation}).
The participants ranked the visibility of the guidance quite highly and commented that it was easily distinguishable from other visual encodings (5.42, \textbf{G3}--\textit{Visible: the guidance and its status are easy to identify and well integrated with the rest of the visual environment}). Furthermore, the participants positively ranked the proposed guidance to use appropriate and expressive encodings (5.25, \textbf{G6}--\textit{Expressive: the visual encoding of guidance is appropriate}), as well as the timeliness of the provided guidance (5.08, \textbf{G7}--\textit{Timely: the guidance is provided on time when needed}). Although expected (as we did not design specifically the guidance to address this), some aspects received lower ratings, providing directions for future work. Most notably, the capability of the guidance to interactively and dynamically adapt to the user's preferences (4, \textbf{G2}--\textit{Adaptive: the guidance system considers the user’s preferences habits and the current task requirements for generating suggestions}). The ability to steer the guidance and evaluate alternative paths of analysis was also ranked low. This implies that presenting alternative paths of analysis or providing provenance of interactions performed and being able to revert the guidance to previous states are aspects that we can improve upon (4.22, \textbf{G4}--\textit{Controllable: the guidance can be steered, allow the user to evaluate alternatives, switch on/off, revert the guidance}). 
Furthermore, participants did not feel like the guidance was self-explanatory and understandable, and this is also reflected in the score (4.82, \textbf{G5}--\textit{Explainable: the guidance and the way it was generated can be easily understood}). Investigating how to improve the explainability of the provided guidance in our proposed approach, and more generally, to other application domains and use cases presents a pressing direction for future research. Finally, based on the outcome of our evaluation we identified that we can improve the capability of the proposed guidance features in providing relevant suggestions depending on the context and recovery from error states (4.62, \textbf{G8}--\textit{Relevant: the guidance helps prevent or recover from errors guides towards a goal helps discover the unexpected provides multiple options and perspectives}).

Summarizing the feedback and results from the questionnaire, we arrive at a final score of \textbf{5.11} which we consider to be a successful evaluation of \sys.

% \textbf{T1} - \textit{Overview.} The system should provide an overview of the temporal information at a glance. Providing an overview is typically the first necessary step in any VA process. This task considers the entire graph as an entity of the analysis~\cite{ahn2014} and is related to the orthogonal ``time'' flattening operation~\cite{17Bach}.

% \textbf{T2} - \textit{Tracking Events.} Understanding the temporal dynamics and the events' frequency helps the user isolate interesting occurrences in time. Events also cause the nodes' trajectories to bend, i.e., make the node change direction. Understanding the \textit{shape}~\cite{ahn2014} of changes in nodes' movement over time (e.g., speed, repetition, etc.) would provide further insights during the exploration of the data. 

% \textbf{T3} - \textit{Investigate relationships.} Each edge event occurrence perturbs the trajectories. Identifying which relationships have the most impact or how often they occur might help the user explain the formation of clusters, or, in general, the phenomenon at hand. This task aligns with the ``point extraction'' task described in Bach et al.'s taxonomy~\cite{17Bach} since we are selecting an individual node and observing how it behaves over time (i.e., changing neighborhood over time).
\section{Discussion}
\label{sec:discussion}
The feedback and comments from the evaluation of our proposed approach gathered by expert interviews offer valuable insights into its usability, effectiveness, and potential improvements. 

Participants found the concept of flattening the space-time cube and the \sys metaphor intuitive, aiding them in comprehending temporal graph data (\textbf{T1}). They effectively performed open-ended exploration tasks (\textbf{T2 \& T3}), generating and answering data-driven questions and reasoning about the network's behavior over time. The timeline and area charts were frequently used to select specific intervals of interest (``when'') and the density maps highlighted areas in the network that seemed intriguing for further inspection. The movement scores and node coloring were appreciated for identify specific entities of the network (``where'') to investigate.

The proposed guidance features (i.e., movement scores, node coloring, and the highlighted timeline intervals) were instrumental in identifying elements of the network and time intervals for in-depth analysis. These features supported \textbf{T2 \& T3} by assisting in understanding temporal dynamics and event frequency. Participants appreciated the ability to control and toggle the elements on or off and explore the data from different perspectives. Finally, the mouseover interactions for exploring the immediate ego network of a specific node and the edge's age encoding (\textbf{T3}) were essential in understanding connectivity and development over time. These interactions facilitated the identification of impactful edges and their frequency, supporting the identification of cluster formation and reasoning about behavior.

\noindent\textbf{Limitations:} 
% Despite \sys's strengths, participants encountered challenges in interpreting certain visualization elements, particularly regarding the layout algorithm and movement score calculations. 
% Due to the ambiguous nature of the mobility scores, it was challenging to reason about the characteristics of high-mobility nodes compared to low-mobility ones. 
The lack of clarity about node movement and layout decisions impacted the interpretability and trust in the guidance features and analysis. Enhancing the explainability and interpretability of the underlying layout algorithm is crucial for improving confidence in the results and interpretation as well as the overall usability of our approach. Furthermore, some scalability concerns emerged with larger datasets, suggesting the need for performance optimization as well as investigating different layout algorithms for a better user experience. Specifically, when depicting networks that have a large number of node trajectories ($>$100) over extended time periods, the visualization's capability of conveying more subtle changes diminished, however, this was also remedied to some extent by narrowing down the interval selection and the resampling frequency. Overlapping node trajectories further complicated the analysis, indicating a need for improved layout algorithms to ensure trajectory separability (e.g., Fig.~\ref{fig:case_study_1}, all nodes’ trajectories are clearly separated this may be a specific property for some of the datasets). 

\section{Conclusion and Future Work} 
In this paper, we presented and described \sys, a guidance-enhanced VA approach to support the analysis and exploration of temporal networks embedded in a space-time cube. 
%Our approach is inspired by the concept of transfer functions in volume rendering. 
%By projecting light down the temporal axis, we visualize different temporal features, such as the nodes' age and persistence, obtaining a flattened 2D visualization of the temporal network, that can be explored by interacting with the timeline. 
We augmented the visualization approach with guidance to better support users in the visual analysis of the data. 
%By providing multiple types of guidance we enhance the exploration of events, significant temporal intervals, interesting nodes, and their relationships. 
We demonstrated the effectiveness of our approach with two use cases and a heuristic evaluation conducted with experts. The case studies depict a scenario where the aim is to explore and extract insights from temporal networks. The results of the expert evaluation highlight the strengths of \sys and outline areas for improvement and refinement. While certain aspects of our interactions and encodings can be improved, the results also demonstrate the benefits of visualizing temporal network data using \sys and providing guidance features to highlight ``when'' and ``where'' there are interesting data points for further inspection. 
Addressing these findings in future iterations of our approach will enhance the scalability for larger datasets, interpretability of the movement score, and the confidence that participants have in their analysis results.
% Overall, the outcomes and comments indicate that there is potential in visualizing temporal graphs by projecting their embedding from the space-time-cube into $2D$. 
The potential of visualization in guiding and supporting the exploration and analysis of temporal graphs is an opportunity for the entire graph drawing and visualization communities.
% Future work should primarily be oriented on conducting a formal evaluation of the method, both from a visual quality perspective, i.e., through a selection of metrics that capture the readability, scalability, and expressiveness of the visualization, and from a user perspective, also assessing the impact of the guidance features included in the system, extending experimentation on other real datasets (e.g.,~\cite{archambault2013map,realitymining}).
% Further work includes investigating how TimeLighting supports users when sampling and identifying temporal network features and patterns. 
%comparison studies between timesliced and event-based visualizations,
%and extending experimentation on other real datasets (e.g.,~\cite{realitymining,archambault2013map}).
%and the potential use of 3D interfaces for a more immersive exploration of space-time cube representations. 
%Furthermore, we plan on conducting a formal evaluation focusing on assessing the quality of the results and insights that can be extracted from our approach as well as determining the extent to which it can scale. 

%% if specified like this the section will be committed in review mode
\ifCLASSOPTIONcompsoc
  % The Computer Society usually uses the plural form
  \section*{Acknowledgments}
\else
  % regular IEEE prefers the singular form
  \section*{Acknowledgment}
\fi
The authors have applied a Creative Commons Attribution (CC-BY) license to any Author Accepted Manuscript version arising from this submission. This work was supported by WWTF [10.47379/ICT19047], FFG DoRIAH [\#880883], FWF ArtVis [10.55776/P35767] and SANE [10.55776/I6635]. The authors acknowledge TU Wien Bibliothek for financial support through its Open Access Funding Programme.
%Our paper was selected by the GD program chairs and invited for publication in TVGC.

%\bibliographystyle{abbrv}
% \bibliographystyle{abbrv-doi}
%\bibliographystyle{abbrv-doi-narrow}
%\bibliographystyle{abbrv-doi-hyperref}
%\bibliographystyle{abbrv-doi-hyperref-narrow}

% references section

\bibliographystyle{IEEEtran}
% argument is your BibTeX string definitions and bibliography database(s)
\bibliography{IEEEabrv,references}
%% VSPACE AND ENLARGE USED TO FIX THE IEEE TEMPLATE BECAUSE ITS BAD
% biographies here
% \vspace{-15 mm}

\newpage
\begin{IEEEbiography}[{\includegraphics[width=1in,height=1.25in,clip,keepaspectratio]{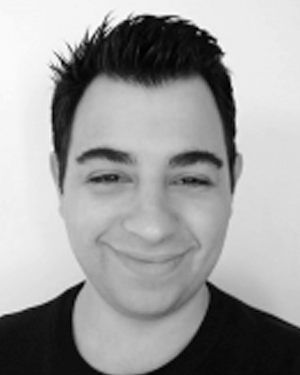}}]{Velitchko Filipov} 
is a post-doc researcher in the research group (CVAST) at the Institute
of Visual Computing and Human-Centered Technology, TU Wien. 
He received his Ph.D. degree from TU Wien, Austria, in 2024 defending his thesis called ``Networks in Time and Space: Visual Analytics of Dynamic Network Representations''. His research interests include visual analytics of
dynamic graphs and networks with a focus on novel representations and interactions.
%Contact him at velitchko.filipov@tuwien.ac.at.
\end{IEEEbiography}

\vspace{-30 mm}
\begin{IEEEbiography}[{\includegraphics[width=1in,height=1.25in,clip,keepaspectratio]{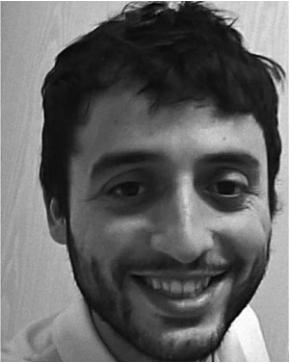}}]{Davide Ceneda} received the PhD degree from TU
Wien, Austria, in 2020 defending a thesis called
``Guidance-Enriched Visual Analytics''. He is a post-doc with the CVAST group at the Institute of Visual Computing and Human-Centered Technology, TU
Wien. His research interests include visualization,
human perception, visual analytics and guidance.
\end{IEEEbiography}
\vspace{-30 mm}
% \newpage
\begin{IEEEbiography}[{\includegraphics[width=1in,height=1.25in,clip,keepaspectratio]{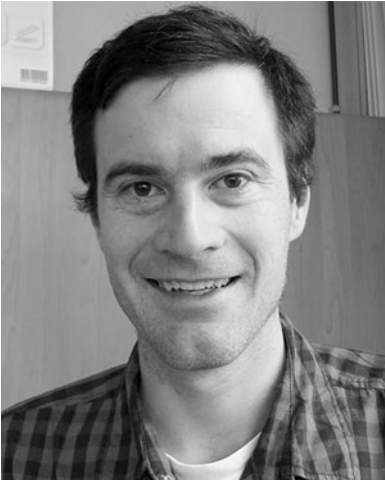}}]{Daniel Archambault} received the PhD degree
from the University of British Columbia, in 2008
and is now a professor of Visualization/Data Science at Newcastle University where he co-leads the scalable computing group.
A main theme of his research is network visualization and evaluating the perceptual effectiveness of such approaches.
\end{IEEEbiography}
\vspace{-30 mm}
\begin{IEEEbiography}[{\includegraphics[width=1in,height=1.25in,clip,keepaspectratio]{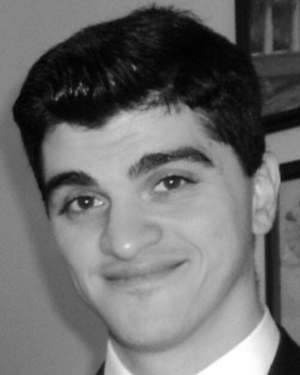}}]{Alessio Arleo}
is an Assistant Professor at the Eindhoven University of Technology, The Netherlands. He obtained a PhD in 2018 at the University of Perugia, Italy,  investigating distributed computing platforms applied to graph drawing. He has been a Post-Doc in the Visual Analytics research unit (CVAST) at the Institute of Visual Computing and Human-Centered Technology, at TU Wien. 
%. He holds a BSc in Computer and Electronic engineering and MSc in Computer and TLC engineering. 
His research interests include temporal network visualization, information diffusion, and distributed graph algorithms.
\end{IEEEbiography}
% \enlargethispage{-13cm} 
\end{document}